\documentclass[times,twocolumn,final,10pt]{cta-author}
{}
{}
{}
\RequirePackage{lineno}
\usepackage{algpseudocode}
\usepackage{algorithm,algorithmicx}
\usepackage{subcaption}
\usepackage{multicol}
\usepackage{booktabs}
\usepackage{multirow,fixltx2e}
\usepackage{multicol}
\usepackage{hhline} 
\usepackage{float}
\usepackage{url} 
\usepackage{textcomp} 
\usepackage{hyperref}
\usepackage{balance}

\usepackage{romannum}
\definecolor{myblue}{rgb}{0, 0, 0}
\definecolor{mygreen}{rgb}{0, 0, 0} 
\definecolor{myviolet}{rgb}{0, 0, 0}
\definecolor{myred}{rgb}{0, 0, 0}
\definecolor{myrosewood}{rgb}{0, 0, 0} 
\begin{document}
\supertitle{Research Article}

\title{CSIS: compressed sensing-based enhanced-embedding capacity image steganography scheme}

\author{\au{Rohit Agrawal$^{1}$}, \au{Kapil Ahuja$^{1\corr}$}}

\address{\add{1}{Mathematics of Data Science (MODS) Laboratory, Indian Institute of Technology Indore, Simrol, Indore, India}
\email{kahuja@iiti.ac.in}}

\begin{abstract}
Image steganography plays a vital role in securing secret data by embedding it in the cover images. Usually, these images are communicated in a compressed format. Existing techniques achieve this but have low embedding capacity. Enhancing this capacity causes a deterioration in the visual quality of the stego-image. Hence, our goal here is to enhance the embedding capacity while preserving the visual quality of the stego-image. We also intend to ensure that our scheme is resistant to steganalysis attacks. 

\par
This paper proposes a Compressed Sensing Image Steganography (CSIS) scheme to achieve our goal while embedding binary data in images. The novelty of our scheme is the combination of three components in attaining the above-listed goals. \textit{First}, we use compressed sensing to sparsify cover image block-wise, obtain its linear measurements, and then uniquely select permissible measurements. Further, before embedding the secret data, we encrypt it using the Data Encryption Standard (DES) algorithm, and finally, we embed two bits of encrypted data into each permissible measurement. This is the first attempt to rigorously embed more than one bit. \textit{Second}, we propose a novel data extraction technique, which is lossless and completely recovers our secret data. \textit{Third}, for the reconstruction of the stego-image, we use the least absolute shrinkage and selection operator (LASSO) for the resultant optimization problem. This has the advantages of fast convergence and easy implementation. This component is also new. 

\par
We perform experiments on several standard grayscale images and a color image, and evaluate embedding capacity, Peak Signal-to-Noise Ratio (PSNR) value, mean Structural Similarity (SSIM) index, Normalized Cross-Correlation (NCC) coefficients, and entropy. {\color{myblue}We achieve 1.53 times more embedding capacity as compared to the most recent scheme. We obtain an average of 37.92 dB PSNR value, and average values close to $1$ for both the mean SSIM index and the NCC coefficients, which are considered good. Moreover, the entropy of cover images and their corresponding stego-images are nearly the same.} {\color{myrosewood}These assessment metrics show that CSIS substantially outperforms existing similar steganography schemes.} 

\end{abstract}

\maketitle

\section{Introduction}\label{introduction}
The primary concern during the transmission of digital data over communication media is that anybody can access this data. Hence, to protect this data from being accessed by illegitimate users, the sender must employ some security mechanisms. In general, there are two main approaches used to protect secret data; cryptography \cite{Stalling} and steganography \cite{Abbas}. In cryptography, the encryption process transforms the secret data, known as plain-text, into cipher-text using an encryption key. This text is in unreadable form, hence, it attracts the opponents to exploit the content of the cipher-text by employing some brute-force attacks \cite{Stalling}. However, steganography avoids this scenario.

Steganography is derived from the Greek words steganos means ``covered or secret" and graphie means ``writing". In steganography, the secret data is hidden into some other unsuspected cover media so that it is visually imperceptible. Here, both the secret data as well as the cover media may be text or multimedia. The media obtain after embedding secret data into cover media is called stego-media. {\color{myblue} Some recent steganography schemes that use text as cover media are \cite{MMT_2020} and \cite{Mathematics_2020}. 
In \cite{MMT_2020}, the authors have proposed an Arabic text steganography scheme, where the secret message is hidden within the text by using Unicode standard encoding. In \cite{Mathematics_2020}, the authors have proposed a character-level text generation-based linguistic steganography scheme, where the secret message is embedded in the text\textquotesingle s content.
}

{\color{myrosewood}Recently, the steganography schemes that use images as the cover media have gained a lot of research interest due to their heavy use in Internet-based applications} Typically, these images are transmitted in a compressed format. So here, we focus on compressed domain-based image steganography. In this, the challenges are; 
\vspace{-5mm}
{\color{myblue}
\begin{enumerate}
    \item Improving the embedding capacity.
    \item Maintaining the quality of the stego-image.
    \item The scheme should be resistant to steganographic attacks.
\end{enumerate}
\vspace{-5mm}
Although images can be embedded into images, our focus is on embedding binary data into images.
}

In the following paragraphs, first we discuss the way in which secret data can be embedded into cover images, then we summarize some existing schemes and their limitations, and finally we argue how the scheme presented in this paper outperforms the existing schemes.

Secret data can be embedded in images by two ways; spatially and by using a transform. In the spatial domain based image steganography scheme, secret data is embedded directly into the image by some modification in the values of the image pixels. {\color{myblue}Some well-known schemes here are listed in \cite{Abbas, Chan1, Steg_2019_FGT, Intech_2016, Intech2_2016, IEEE_2016, JIS_2018, Arabian_3D_Mesh_2018}.} In the transform domain based image steganography scheme, first, the image is transformed into frequency components, and then the secret data is embedded into these components. {\color{myblue}Some commonly used such schemes are JSteg \cite{Jsteg}, F5 \cite{F5}, and Outguess \cite{Outguess}. Some other techniques, which do not carry specific names are given in references \cite{Chang, Liu, Pan, Steg_ExpertSystem, Steg_Rohit, JAIS_2017, Elsvier_2018, IEEE_2019, Rohit_arXiv}.
} 

{\color{myviolet}The spatial domain based image steganography outperforms the transform domain one in terms of embedding capacity, but the stego-image has a high amount of redundant data. Digital images transmitted through communication media are usually of this type. Since transform based schemes reduce the redundancy present in the image and represent it in a compressed form, they are preferred for transmission.} 

Most of the transform domain based scheme follow either Discrete Cosine Transform (DCT) or Wavelet Transform (WT). The DCT based schemes are also called the JPEG compression based image steganography techniques. {\color{myblue}Several variants of DCT based schemes have been proposed in the literature \cite{Jsteg, F5, Outguess, Chang, Liu, Pan, Steg_Rohit, JAIS_2017, Elsvier_2018, IEEE_2019, Rohit_arXiv}. For the schemes \cite{Jsteg, F5, Outguess, Chang, Liu, Pan, Steg_Rohit, Elsvier_2018, IEEE_2019}, secret data is binary bits, and for \cite{JAIS_2017, Rohit_arXiv}, secret data is images.}

In \cite{Jsteg, Outguess, Chang}, the secret data is embedded by flipping the least significant bit (LSB) of the quantized DCT coefficients obtained from the cover image. This process is considered as a direct embedding mechanism. {\color{myblue}Alternatively, methods in \cite{F5, Liu, Pan, Steg_Rohit, Elsvier_2018, IEEE_2019, JAIS_2017, Rohit_arXiv} are considered as indirect steganography schemes in which the quantized DCT coefficient values are altered according to certain secret message bits or secret image pixels.} {\color{myblue}By steganalysis, which is the study of detecting the secret data hidden using steganography, it has been observed that the indirect steganography mechanism is superior to the direct one due to its capability in resisting certain statistical attacks. The most common statistical attacks are the chi-square test, and the shrinkage effect \cite{Westfeld, Fridrich, Patsakis}.} {\color{myblue}Hence, the schemes \cite{Jsteg, Outguess, Chang} are not resistant to such attacks, while the schemes \cite{F5, Liu, Pan, Steg_Rohit, Elsvier_2018, IEEE_2019, JAIS_2017} are resistant to them, but their embedding capacity is limited. {\color{myrosewood}If we try to increase the embedding capacity of the later schemes, then the quality of the stego-images gets degraded.} The scheme \cite{Rohit_arXiv} has high embedding capacity with resistance to steganographic attacks, but here, the secret data is the images, which is different from our goal of embedding binary data in images.} 

Most recent Wavelet transform based steganography schemes are given in \cite{Steg_ExpertSystem, DWT_2019}. In \cite{Steg_ExpertSystem}, the authors have proposed a steganography scheme based upon edge identification and XOR coding that uses Wavelet transformation. 
This scheme is resistant to steganographic attacks, but here also the embedding capacity is significantly less. As above, if we try to increase embedding capacity, then the quality of stego-image gets degraded. {\color{myrosewood}The scheme given in \cite{DWT_2019} embeds a medical image into a cover image using Redundant Integer Wavelet Transform (RIWT) and DCT.} This scheme\textquotesingle s purpose is again different from ours of embedding binary data in images.

As discussed above, conventional transform domain based image steganography schemes provide good visual quality stego-image and are resistant to steganographic attacks, but their embedding capacity is limited. If we try to increase their embedding capacity, then the stego-image quality degrades. To overcome this limitation, in this manuscript, we utilize another paradigm, the compressed sensing, which also fulfills all the requirements of image steganography. Next, we present literature regarding compressed sensing-based steganography schemes. These works help to achieve some of the above objectives of steganography but not all, which we do. 

In \cite{Sreedhanya}, and \cite{Rohit}, steganography schemes based on compressed sensing and Singular Value Decomposition (SVD) have been presented. In these schemes, secret medical image data is embedded into an image cover media. Both these approaches use a similar embedding approach, but use compressed sensing differently. In these, first, encrypted measurements of the secret image are obtained using the compressed sensing technique, and then these encrypted measurements are embedded into the cover image using SVD based embedding algorithm. In \cite{Sreedhanya}, the PSNR (Peak Signal-to-Noise Ratio, discussed in Section \ref{PSNR}) value of the stego-image is greater than 30 dB, which shows that it produces good quality stego-images. {\color{myblue}But the PSNR value of the constructed secret image is very low, i.e. the quality of the secret image is degraded very much.} In contrast, in \cite{Rohit}, both the stego-image as well as the reconstructed secret image preserved good visual quality. But, the goal in both these schemes is different from ours. In these schemes, the secret data is an image. {\color{myblue}If these techniques are applied on binary data that we want to embed, the information will be lost.} In \cite{Pan}, the authors have proposed an image steganography scheme based on sub-sampling and compressed sensing. In this scheme, the PSNR value of the stego-image is greater than 30 dB, also the secret data is binary. However, the embedding capacity in this scheme is very low. 

{\color{myblue}Moreover, some other compressed sensing-based image steganography schemes are listed in \cite{Patsakis}, \cite{ Patsakis1}, and \cite{MMT_2019}. In \cite{Patsakis}, the authors have presented the application of compressed sensing to detect steganographic content in the LSB steganography scheme.} In \cite{Patsakis1}, the authors have proposed a DCT steganography classifier based on a compressed sensing technique. Here, the original image is identified from a set of images containing the original image and some instances of stego images. {\color{myblue}In \cite{MMT_2019}, the authors have proposed an image steganalysis technique for secret signal recovery}. These steganography schemes are not related to our work because the focus of \cite{Patsakis} and \cite{MMT_2019} is steganalysis, while \cite{ Patsakis1} focuses on steganography classifier. Hence, we do not discuss these schemes in detail. 

{\color{myblue}The scheme that we propose satisfies all the goals mentioned in the earlier paragraphs, i.e. increased embedding capacity without degrading the quality of stego-images as well as making the scheme resistant to steganalysis attacks. Our scheme has three components, which we discussed next. The \textit{first} component of our scheme consists of three parts; (\romannum{1}) we use compressed sensing to sparsify cover image block-wise and obtain linear measurements. Here, we design an adaptive measurement matrix instead of using a random one. Using our adaptive measurement matrix, we uniquely select a large number of permissible measurements compared to existing schemes. Hence, we achieve a high embedding capacity. Moreover,  these measurements act as encoded transformed coefficients, and hence, this adds security to our proposed scheme as well; (\romannum{2}) we encrypt the secret data using the Data Encryption Standard (DES) algorithm \cite{Stalling}. This adds another layer of security to our scheme; (\romannum{3}) we embed two bits of secret data into each permissible measurement instead of commonly embedding one bit per measurement. This is a first attemp to rigorously embed more than one bit. \textit{Second}, we completely extract secret data without any loss using our extraction algorithm. \textit{Third}, we use the alternating direction method of multipliers (ADMM) solution of the least absolute shrinkage and selection operator (LASSO) formulation of the underlined optimization problem in the stego-image construction. The advantages of using ADMM and LASSO are that they have broad applicability in the domain of image processing, require a little assumption on the objective function\textquotesingle s property, have fast convergence, and are easy to implement. This is also a completely new contribution.}

For performance evaluation, we perform experiments on standard test images. To check the quality of stego-image, we reconstruct it from the obtained modified measurements and then compare it with its corresponding cover image. {\color{myblue}We evaluate embedding capacity, Peak Signal-to-Noise Ratio (PSNR) value, mean Structural Similarity (SSIM) index, Normalized Cross-Correlation (NCC) coefficient, and entropy. We achieve 1.53 times more embedding capacity when compared with the most recent scheme of this category. We achieve a maximum of 40.86 dB and an average of 37.92 dB PSNR values, which are considered good. The average values of mean SSIM index and NCC coefficients are close to 1, which are again considered good. Moreover, the entropy of cover images and their corresponding stego-images are nearly the same.} In the Experimental Results section, we also show that our scheme outperforms existing compression based steganography schemes \cite{Jsteg, F5, Outguess, Liu, Pan, Steg_Rohit, Steg_2019_FGT, Steg_ExpertSystem}.  

The rest of the paper has four more sections. Section \ref{compressed sensing} describes the compressed sensing technique. Section \ref{propose method} explains our proposed steganography scheme including embedding of the data, extracting it, and stego-image reconstruction process. Section \ref{experimental results} presents the experimental results. Finally, Section \ref{conclusion} gives conclusions and future work.

\section{Compressed Sensing}\label{compressed sensing}
Compressed sensing is used to acquire and reconstruct the signal efficiently. Traditionally, the successful reconstruction of the signal from the measured signal must follow the popular Nyquist/ Shannon sampling theorem, which states that the sampling rate must be at least twice the signal bandwidth. {\color{myblue}In many applications such as image, audio, video, data mining, and wireless communications \& networks, where the signal is sparse or sparsified in some domain, the Nyquist rate is too high to achieve.} There is a fairly new paradigm, called compressed sensing that can represent the sparse signal by using a sampling rate significantly lower than the Nyquist sampling rate \cite{Candes, Donoho}. {\color{myblue}Hence, the application of compressed sensing has gained popularity in many areas. Some of them are image processing \cite{Romberg}, radar system \cite{Steeghs}, MRI Imaging \cite{CS_MRI}, and noise separation from data \cite{Romberg}.} 

Compressed sensing projects the sparse signal onto a small number of linear measurements in such a way that the structure of this signal remains the same. The sparse signal can be reconstructed approximately from these measurements by an optimization technique. {\color{myblue}However, the reconstruction of the signal is possible only when the original signal is sparse, and it satisfies the Restricted Isometric Property (RIP) \cite{Candes2} (discussed in Section \ref{CS sensing matrix})}. {\color{myblue}If the original signal is not sparse, then it can often be artificially sparsified}. A brief description of signal sparsification, obtaining linear measurements, and reconstruction of the approximate sparse signal is given next.

\subsection{Signal Sparsification}\label{CS signal sparsification}
Let the original signal be \textbf{$x\in  \mathbb{R}^{N\times 1}$}. The signal \textit{x} is \textit{K} sparse when it has maximum \textit{K} number of non-zeros coefficients, i.e. $\left \| x \right \|_{0}\leq K$, where $|| \cdot ||_{0}$ denotes the $\ell_{0}-norm$ of a vector, and the remaining coefficients are zero or nearly zero. Let the original signal \textit{x} not be sparse and be represented in-terms of $\left \{ \psi _{i} \right \}_{i=1}^{N}$ basis vectors each of length $N\times 1$, then 
\begin{linenomath*}
\begin{equation}
	s=\Psi^{T} x,
\end{equation}
\end{linenomath*}
where, $s\in  \mathbb{R}^{N\times 1}$ and $\Psi =\left [ \psi _{1}, \psi _{2},...,\psi _{N} \right ]\in  \mathbb{R}^{N\times N}$ is an orthogonal matrix. If $K \ll N$ then this signal is sparsifiable \cite{Candes3}, $s$ is the sparse representation of $x$, and $\Psi$ is the corresponding sparsification matrix.
  
\subsection{Sensing Matrix and Linear Measurements}\label{CS sensing matrix}
In the compressed sensing framework, we acquire $M (M<N)$ linear measurements from the inner product between the original signal $x\in  \mathbb{R}^{N\times 1}$ and \textit{M} measurement vectors $\left \{ \phi _{i} \right \}_{i=1}^{M}$, where $\phi _{i}\in  \mathbb{R}^{N\times 1}$. Considering the measurement/ sensing matrix as $\Phi =\left [ \phi _{1}^{T}; \phi _{2}^{T};...;\phi _{M}^{T} \right ]\in  \mathbb{R}^{M\times N}$, the measurements $y\in  \mathbb{R}^{M\times1}$ are given as \cite{Candes3}
\begin{linenomath*}
\begin{equation}
y=\Phi x.
\end{equation}
\end{linenomath*}
If the input signal is not sparse but sparsifiable, then using the above theory we get 
\begin{linenomath*}
\begin{equation}
y=\Phi \Psi s=\Theta s,
\end{equation}
\end{linenomath*}
{\color{myblue}where $\Theta=\Phi \Psi$ is again the measurement matrix of size $M\times N$. Usually, in the compressed sensing framework, the measurement matrix is nonadaptive. That is, the measurement matrix is fixed and does not depend on the signal.} However, in certain cases, adaptive measurements can lead to significant performance improvement. 

The main concern here is to design the measurement matrix in such a way so that the most of the information and the structure of the signal is preserved in the measurements. This would imply that original signal would be recovered efficiently from these measurements. {\color{myblue}To achieve this, for all \textit{K-sparse} signals $s$, the measurement matrix should hold the following inequality  \cite{Candes2}.
\begin{linenomath*}
\begin{equation}
\left ( 1-\delta _{K} \right )\leq \frac{\left \| \Theta s \right \|_{2}^{2}}{\left \| s \right \|_{2}^{2}}\leq \left (1+\delta _{K} \right),
\end{equation}
\end{linenomath*}
where $\delta _{K} \in (0, 1)$ is an isometric constant. The above inequality is called the RIP that informally says that the $\ell 2-norm$ of the sparse signal $s$ and the measurement $\Theta s$ should be comparable.} {\color{myblue}Apart from satisfying the RIP, the minimum number of measurements required, i.e. the minimum value of $ M $, is also a concern in the measurement matrix design.} 

\subsection{Reconstruction of the Approximate Signal}\label{CS Reconstruction}
{\color{myblue}As discussed in the previous subsection, size of the measurement $y=\Phi x=\Phi \Psi s=\Theta s$ is less than the size of the original signal $s$. Hence, the reconstruction of the signal from measurements becomes an ill-posed problem. That is, the solution of an under-determined linear system of equations is to be found.} If the matrix $\Theta$ satisfies the RIP, then the sparse signal $s$ can be reconstructed approximately by solving the following optimization problem \cite{Baraniuk}:
\begin{linenomath*}
\begin{equation}
\begin{aligned}
&\min_s \left \{\text{number of i such that} \: s\left ( i \right )\neq 0 \right \} \\ 
&\text{Subject to} \hspace{0.1cm} \Phi \Psi s=y.
\end{aligned}
\end{equation}
\end{linenomath*}

In the above equation, the function to be minimized is simply the number of nonzero coefficients in the vector $s$. This equation is referred to as $\ell_{0}-norm$ minimization problem. It is combinatorial and an NP-hard problem \cite{Baraniuk}. The other approach is to substitute the $\ell_{0}-norm$ by the closest convex norm, i.e. the $\ell_{1}-norm$, or 
\begin{linenomath*}
\begin{equation}
\begin{aligned}\label{eq: CS reconstruction l1}
& \min_s \left \| s \right \|_{1} \\ 
& \text{Subject to} \hspace{0.1cm} \Phi \Psi s=y,
\end{aligned}
\end{equation}
\end{linenomath*}
{\color{myblue}where $|| \cdot ||_{1}$ denotes the $\ell_{1}-norm$ of a vector.} The approach to reconstruct the sparse signal $s$ by solving the above equation is termed as a convex optimization method. 

{\color{mygreen}Other approaches such as Greedy based (OMP \cite{OMP}, CoSaMP \cite{Needell}), sparse reconstruction by separable approximation \cite{Stephen}, Bayesian strategy \cite{David}, and ADMM solution of the LASSO formulation of the above optimization problem can also be used to reconstruct the sparse signal from the measurements \cite{ADMMSBoyd, Lasso}.


Next, we give a brief idea of LASSO and ADMM, which we use. The general LASSO problem is given as \cite{Lasso}
\begin{equation}\label{eq:Lasso}
\min_z \left\{\frac{1}{2}\Vert A z -b \Vert_{2}^{2}+\lambda\Vert z \Vert_{1}\right\},
\end{equation}
where $z\in \mathbb{R}^n$, $A\in \mathbb{R}^{p\times n}$, $b\in \mathbb{R}^p$, $\Vert \cdot \Vert_{2}$ is the $\ell_2$ norm and $\lambda>0$ is a scalar regularization parameter also called Lagrangian parameter \cite{ParaLarPD}. Further, \eqref{eq:Lasso} is transformed into a form solvable by ADMM \cite{ADMMSBoyd}. That is
\begin{align}\label{eq:Lasso_ADMM}
\begin{split}
& \min\limits_{z, z_1} \left\{\frac{1}{2}\Vert A z -b \Vert_{2}^{2} + \lambda\Vert z_1 \Vert_{1}\right\} \\ 
& \text{ Subject to } z - z_1 = 0.
\end{split}
\end{align}
Finally, ADMM solve the above optimization problem. 

Now, we discuss how to solve our signal reconstruction problem, i.e. \eqref{eq: CS reconstruction l1} by LASSO and ADMM. For our case, $\Theta = \Phi \Psi $ is the measurement matrix, and $\Theta \in \mathbb{R}^{M\times N}$. In the compressed sensing framework, matrix $\Theta$ is underdetermined, i.e. $M < N$. Hence, there is equivalent solution of \eqref{eq: CS reconstruction l1}, which is given as \cite{l1_Lasso}
\begin{linenomath*}
\begin{equation}
\begin{aligned}\label{eq: CS reconstruction Lasso}
& \min_s \left\{ \frac{1}{2} \left \| \Theta s - y \right \|_{2}^2 + \lambda \left \| s \right \|_{1} \right\} \\ 
\end{aligned}
\end{equation}
\end{linenomath*}
Here, we observe that \eqref{eq: CS reconstruction Lasso} is equivalent to \eqref{eq:Lasso} with $\Theta = A$, $s=z$ and $y=b$.
}

Finaly, we briefly mention a theoretical result related to reconstruction. In \cite{Candes4}, it is shown that for sufficiently small constant \textit{C ($C>0$)}, the \textit{K-sparse} signal $s$ of size \textit{N} can be approximately reconstructed from \textit{M} measurements $y$ if $M\geq CK\left ( \log N \right )$. After recovering the sparse signal $s$, the original signal $x$ can be obtained as $x=\Psi s$. For us, this property holds.

\section{Proposed Method}\label{propose method}
Our proposed compressed sensing-based image steganography scheme consists of the following components; data embedding, data extraction, and stego-image construction, which are discussed in the respective sections below.

{
		\begin{algorithm}[!h]
\scriptsize 
		\caption{Embedding Rule}\label{alg:embedding rule}
		\begin{algorithmic}[1]
			\renewcommand{\algorithmicrequire}{\textbf{Input:} }
			\renewcommand{\algorithmicensure}{\textbf{Output:} }
			\Require \quad
			\begin{itemize}
				\item $y$: Sequence of transform coefficients.
				\item S: Encrypted secret bit sequences which is to be embedded.
			\end{itemize} 
			\Ensure \quad
			\begin{itemize}
				\item $z$: The modified version of transform coefficients. 
			\end{itemize}
			\If {($length(S) < 2\times length(y)$)}
			\For {$j = 1$ to $length(y)$}
			\If {(${y}\left ( j \right )= -1$ or ${y}\left ( j \right )= 0$ or ${y}\left ( j \right )= +1$)}
			\State ${z}={y}$	\quad (Do not embed in these measurements)
			\Else
			\If {(${y}\left ( j \right )\%2= 0$)}
			\If {(${y}\left ( j \right )\%4= 0$)}
			\If {($S\left ( j \right )= 00$)}
			\State ${z}={y}+1$
			\ElsIf {($S\left ( j \right )= 01$)}
			\State ${z}={y}$
			\ElsIf {($S\left ( j \right )= 10$)}
			\State ${z}={y}-1$
			\ElsIf {($S\left ( j \right )= 11$)}
			\State  ${z}={y}+2$ or ${z}={y}-2$
			\EndIf
			\Else
			\If {($S\left ( j \right )= 00$)}
				\If {(${y}\neq 2$)}
					\State ${z}={y}-1$
					\Else 
					\State ${z}={y}+3$
				\EndIf	
			\ElsIf {($S\left ( j \right )= 01$)}
				\If {(${y}\neq -2$)}
					\State ${z}={y}+2$
					\Else 
					\State ${z}={y}-2$
				\EndIf	
			\ElsIf {($S\left ( j \right )= 10$)}
				\If {(${y}\neq -2$)}
					\State ${z}={y}+1$
					\Else 
					\State ${z}={y}-3$
				\EndIf	
			\ElsIf {($S\left ( j \right )= 11$)}
			\State ${z}={y}$
			\EndIf
			\EndIf
			\Else
			
			\If {\big((${y}\left ( j \right )-1)\%4= 0$\big)}
			\If {($S\left ( j \right )= 00$)}
			\State ${z}={y}$
			\ElsIf {($S\left ( j \right )= 01$)}
			\State ${z}={y}-1$
			\ElsIf {($S\left ( j \right )= 10$)}
			\State ${z}={y}-2$
			\ElsIf {($S\left ( j \right )= 11$)}
			\State ${z}={y}+1$
			\EndIf
			\Else
			\If {($S\left ( j \right )= 00$)}
			\State ${z}={y}+2$
			\ElsIf {($S\left ( j \right )= 01$)}
			\State ${z}={y}+1$
			\ElsIf {($S\left ( j \right )= 10$)}
			\State ${z}={y}$
			\ElsIf {($S\left ( j \right )= 11$)}
			\State ${z}={y}-1$
			\EndIf
			\EndIf
			\EndIf
			\EndIf
			\EndFor	
			\Else
			\State Whole secret data cannot be embedded. Try short length secret data.  
			\EndIf 	\\
			\Return $z$
\end{algorithmic}
\end{algorithm}
}
\subsection{Data Embedding}\label{sec:data embedding}
	
\begin{figure*}[!h]
	\centering
		\includegraphics[width=0.8\textwidth]{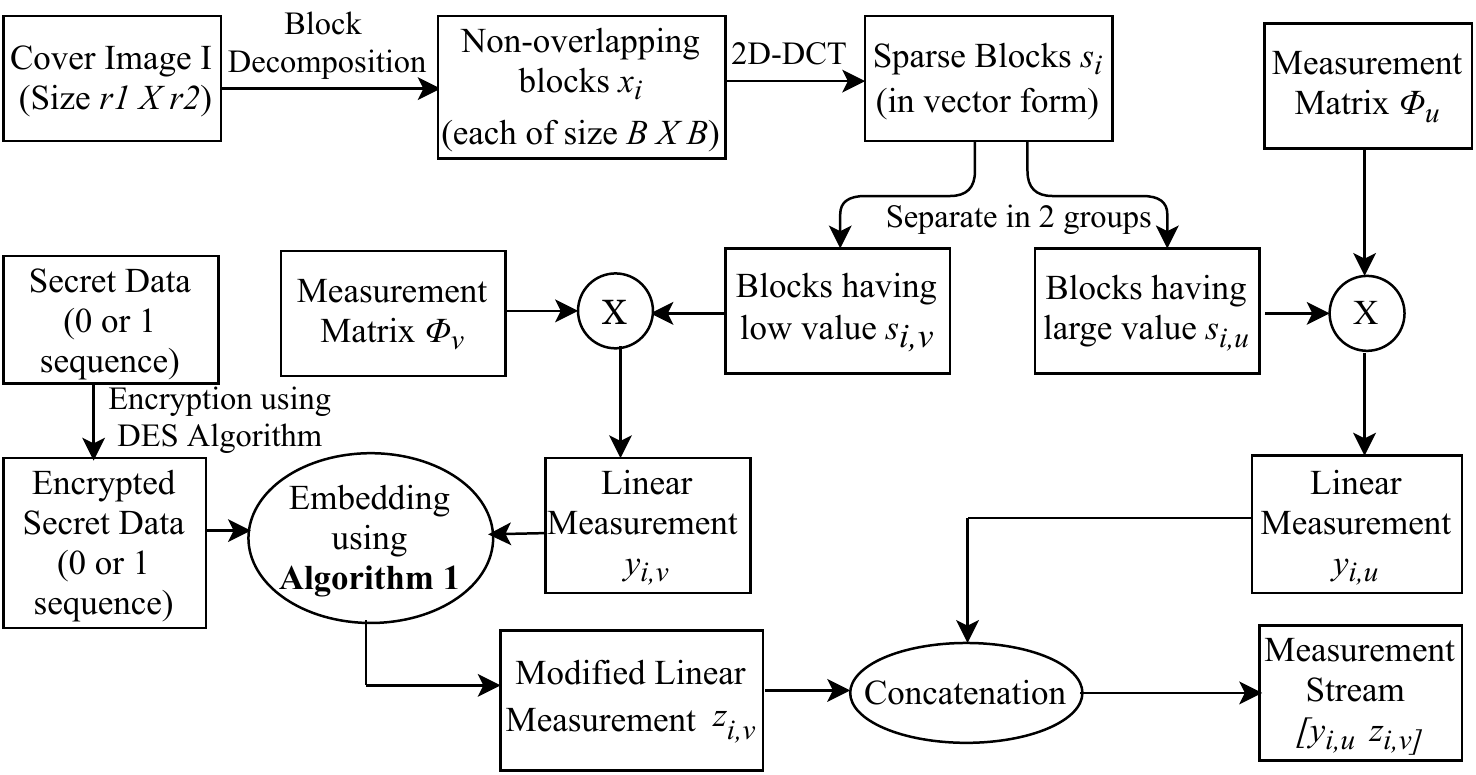}
		\caption{The Embedding Process}
		\label{Figure:Embedding process}
\end{figure*}

The first step in any compressed sensing-based image steganography scheme is the input image\textquotesingle s sparsification if it is not sparse at the start. This step is equivalent to the signal sparsification of Section \ref{CS signal sparsification}. {\color{myblue}Methods such as K-SVD, DCT, Discrete Walsh Transform, Stationary Wavelet Transform, and Discrete Rajan Transform provide good sparsification.} Since the distortion due to DCT is less, we use it as our sparsifying agent. To further reduce the distortion, instead of sparsifying the whole image at once, first, we decompose the cover image into non-overlapping blocks of the same size, and then each block is sparsified. 

{\color{myblue}Let the image I\textquotesingle {s} size be $r1 \times r2$ and each block size be $B\times B$, then we have $(r1\times r2)/B^2$ number of blocks. In our case, $r1=r2$ and $B$ completely divides $r1$.} The block-wise sparsification is now done as
\begin{linenomath*}
	\begin{equation}\label{eq: propose sparsification}
	s_{i}=DCT\left(x_{i}\right), 
	\end{equation}
\end{linenomath*}
where $i=1, 2, \cdots, (r1\times r2)/B^2$, $x_{i}$ and $s_{i}$ are the $i^{th}$ original and sparse blocks of the same size, i.e. $B\times B$, respectively. {\color{myblue}Next, we convert each block into their vector representation by stacking them column-wise. Thus, $s_{i}$ becomes a vector of size $B^2\times 1$.} Because of sparsification, each sparse vector has few coefficients of large values and the remaining coefficients of very small values or zero. Hence, we categories each vector into two groups. Let $p_1$ be the number of coefficients having large values and $p_2$ be the number of coefficients having small values or zero values. Note that here, $p_1<p_2$ as each of these vectors are sparse in nature and $p_1+p_2=B^2$. We represent each vector in two groups based upon these coefficients, i.e. $s_{i,u}\in \mathbb{R}^{p_1}$ and $s_{i,v}\in \mathbb{R}^{p_2}$. Now, we project each sparse vector onto linear measurements using a measurement matrix, which is equivalent to Section \ref{CS sensing matrix}.

There are two ways to choose the measurement matrix: either randomly or deterministically. {\color{myblue}Randomly generated matrices such as the Independent and Identically Distributed (i.i.d.) Gaussian matrix, the Bernoulli matrix or other matrices generated by probabilistic methods are nonadaptive, although they satisfy the RIP.} Deterministically generated matrices are the ones that are designed such that specific properties are satisfied, e.g., adaptiveness and the RIP. We design a deterministic matrix that is adaptive to our sparse vector since this improves the efficiency of compressed sensing. To achieve RIP here, the projected linear measurements are enforced to have almost the same $\ell_2-norm$ as that of the sparse vector. 

One way to design the measurement matrix is to first analyze the distribution of all $B^2$ coefficients in each sparse vector, and then find the $m$ indices out of these that give maximum $\ell_2-norm$ \cite{Xiaorong}. That is ,  
\begin{linenomath*}
\begin{equation}\label{eq: find m indices}
E_{max}^{|m|}=\max_{i\in m\subset B^2}\left \| s_{i} \right \|_{2}^{2},
\end{equation}
\end{linenomath*}
{\color{myblue}where $|m|$ is the number of entries in set $m$ and $E_{max}^{|m|}$ is a variable that stores the maximum value of square of $\ell_2$-norm of vector $s_i$ for $i\in m\subset B^2$.} However, in this paper, we use the property of DCT to design the measurement matrix. This property states that DCT coefficients can be divided into three sets; low frequency, middle frequency, and high frequency components. Low frequency corresponds to the overall image information, middle frequency corresponds to the structure of the image, and high frequency corresponds to the noise or small variance. For image reconstruction, only lower and middle frequency components are useful. {\color{myblue}Hence, we select $m$ indices out of all $B^2$ indices that correspond to these two sets of frequency \cite{Chang}.} Here, $|m|$ is a user-defined parameter such that $p_1 < |m| < p_1+p_2$, and is discussed in Experimental Results section. As discussed earlier, in this subsection we have two groups of sparse vectors $s_{i,u}$ and $s_{i,v}$. Hence, we design two different measurement matrices $\Phi_{u}$ and $\Phi_{v}$ corresponding to $s_{i,u}$ and $s_{i,v}$, respectively. 

Since $\left \| s_{i,u} \right \|_{2}$ is close to $\left \| s_{i} \right \|_{2}$ because $s_{i,u}$ contains large value coefficients of $s_{i}$, we project $s_{i,u}$ onto the same number of linear measurements. Thus, we have $\Phi_{u}=\alpha I_{p_1}$, where $I_{p_1}$ is the identity matrix of size $p_1 \times p_1$, and $\alpha$ is a small constant.

As mentioned in Section \ref{CS sensing matrix}, the main purpose of measurement matrix is to project the sparse vector onto less number of linear measurements. Hence, we project $s_{i,v}$  onto $|m| - p_1$ measurements or the size of $\Phi_{v}$ is $(|m|-p_1)\times p_2$. {\color{myblue}To construct $\Phi_{v}$, we first take a random Hadamard matrix of size $p_2\times p_2$, which is a standard procedure in compressed sensing literature \cite{Hadamard}, and then we choose $|m|-p_1$ rows from the available $p_2$ rows.} These rows map to the last of $|m|-p_1$ indices from the index set $m$. This is because the first $p_1$ indices have the overall image information, and hence, map to construction of $\Phi_u$.

We use the same measurement matrices for all blocks. This is because, for all blocks of an image, the distribution of coefficients of the generated sparse vectors is almost the same. Thus, for each block $i=1,2, \dots, (r1\times r2)/B^2$, {\color{myblue}the block-wise linear measurements vector $y_i\in \mathbb{R}^{|m|}$ is given as}
\begin{linenomath*}
\begin{equation}\label{eq: propose measurements}
y_i=\begin{bmatrix}y_{i,u}\\ y_{i,v} \end{bmatrix}=
\begin{bmatrix}
\Phi _{u}s_{i,u}\\ \Phi _{v}s_{i,v}	
\end{bmatrix}.
\end{equation}
\end{linenomath*}
Using the standard terminology \cite{Candes, Donoho}, the measurements $y_{i,u}$ are called the ordinary samples or non-compressed samples, and the measurements $y_{i,v}$ are called the compressed sensing samples. 

{\color{myblue}Next, we discuss the encryption process of the secret data $D$ that is to be embedded. This data is a sequence of $0s$ and $1s$.} As mentioned in the Introduction, this provides an extra layer of security to the embedded data. For this, we first encrypt this data by using DES algorithm to obtain the encrypted secret data $S$ (which is also a sequence of $0s$ and $1s$) \cite{Stalling}. DES is a fairly standard algorithm used for data encryption \cite{Stalling}. Then, we represent $S$ as a set of two-two bits, i.e. $S=\left\lbrace S_{1}, S_{2}, \dots, S_{n}\right\rbrace$, where each $S_{L}$ consists of two bits.

Next, we embed the secret data in our linear measurements $y_i$. The embedding rule is summarized in \textbf{Algorithm \ref{alg:embedding rule}}, and helps to embed two bits into the transform coefficients. The rule is designed in such a way so that the secret data could be extracted without any loss, discussed in Data Extraction and Experimental Result sections. We embed the data in $y_{i,v}$ and not $y_{i,u}$. This is because $y_{i,u}$ corresponds to sparse vector coefficients of large values, and embedding in it leads to degradation of image quality. Further, in $y_{i,v}$, the secret data is embedded selectively. We do not embed in $y_{i,v}$ with measurement value of $-1$, $0$ and $1$. {\color{myblue}This is because our embedding algorithm concatenates the measurement values with integers from $-3$ to $+3$, and if these values are $-1$, $0$ or $1$, then we may end up getting many $0s$ after concatenation, which leads to difficulty in the extraction process.} After embedding in other measurement values of $y_{i,v}$, we obtain the modified $y_{i,v}$, which is termed as $z_{i,v}$. That is,
\begin{linenomath*}
	\begin{equation}
	z_{i,v}= 
	\begin{cases}
	y_{i,v} & \text{if } y_{i,v}=-1, \: 0 \: or \: 1\\
	y_{i,v}+c             & \text{otherwise},
	\end{cases}
	\end{equation}
\end{linenomath*}	
where $c\in \{-3,-2,-1,0, 1,2,3\}$. We obtain our stego-data by concatenating the measurements $y_{i,u}$ and $z_{i,v}$ as $\begin{bmatrix} y_{i,u} \\ z_{i,v} \end{bmatrix}$. The block diagram for this complete data embedding process is given in Fig. \ref{Figure:Embedding process}. 

\subsection{Data Extraction}\label{sec:data extraction}
In this section, we explain the process of extracting embedded secret data from our stego-data. The steps of this extraction process are given below, which are exactly reverse to our data embedding process.

\begin{enumerate}
	\item Separate the measurements $z_{i,v}$ from the stego-data, i.e. $\begin{bmatrix} y_{i,u} \\ z_{i,v} \end{bmatrix}$, where $i=1,2, \dots,(r1\times r2)/B^2$ is the block number, and $u,v$ are indices available from the previous subsection.
	\item Extract only those measurements from $z_{i,v}$ whose values are not equal to $-1$, $0$ or $1$. The embedding rule ensures that the embedded data could be extracted without loss. In other words, \textbf{Algorithm \ref{alg:embedding rule}} ensures that no secret data is embedded in measurements with values $-1$, $0$ and $1$.
	\item Extract the encrypted message ${S}'$ from the measurements obtained in the above step by applying \textbf{Algorithm \ref{alg:Extraction rule}}.
	\item Decrypt this ${S}'$ by DES algorithm, and obtain the extracted secret data $D'$.
\end{enumerate}

\begin{figure}
	\centering
	\includegraphics[width=0.45\textwidth]{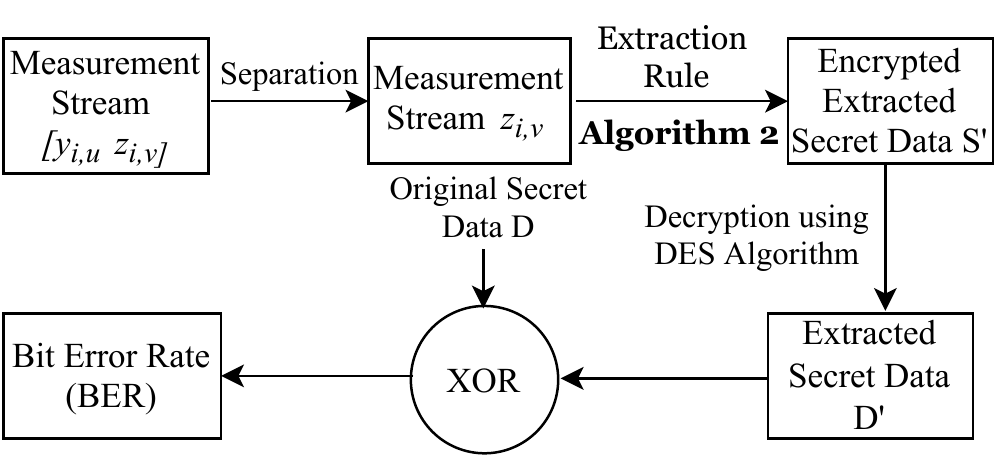}
	\caption{The Extraction Process}
	\label{Figure:Extraction process}
\end{figure}

Now, we check the correctness of this extracted secret data $D'$ by comparing it with original secret data $D$. For this, we use the Bit Error Rate (BER), which is given as \cite{Pan}
\begin{linenomath*}
\begin{alignat}{2}
& \text{Error Bits (EB)}=D \bigoplus {D}',	\\	
& BER=\frac{\text{Number of ones in EB}}{\text{Size of D}}\times 100\%,
\end{alignat}
\end{linenomath*}
where $\bigoplus$ denotes the bitwise XOR/ Exclusive OR operation. The BER value for our steganography scheme is $0\%$, i.e. we successfully extract complete secret data without any error. This is the property of our embedding rule. The above extraction process is represented via a block diagram in Fig. \ref{Figure:Extraction process}.

{
	\begin{algorithm}[t]
	\scriptsize 
		\caption{Extraction Rule}\label{alg:Extraction rule}
		\begin{algorithmic}[1]
			\renewcommand{\algorithmicrequire}{\textbf{Input:}}
			\renewcommand{\algorithmicensure}{\textbf{Output:}}
			\Require \quad
			\begin{itemize}
				\item $z$: Sequence of modified linear measurements. These are $z_{i,v}$ that are not having value equal to $0$, $1$ or $-1$. See extraction process in Section \ref{sec:data extraction}.
			\end{itemize} 
			
			\Ensure \quad
			\begin{itemize}
				\item ${S}'$: Encrypted secret bit sequences.
			\end{itemize}
			\For {$j = 1$ to $length(z)$}
			\If {(${y}\left ( j \right )= -1$ or ${y}\left ( j \right )= 0$ or ${y}\left ( j \right )= +1$)}
			\State Continue 
			\Else
			\If {(${z}\left ( j \right )\%2= 0$)}
			\If {(${z}\left ( j \right )\%4= 0$)}
			\State ${S}'\left ( j \right )= 01$
			\Else
			\State ${S}'\left ( j \right )= 11$
			\EndIf
			\Else
			\If {\big((${z}\left ( j \right )-1)\%4= 0$\big)}
			\State ${S}'\left ( j \right )= 00$		
			\Else 
			\State ${S}'\left ( j \right )= 10$
			\EndIf
			\EndIf
			\EndIf
			\EndFor		\\		
			\Return ${S}'$
		\end{algorithmic}
	\end{algorithm}
}

\subsection{Stego-Image Construction}\label{Stego Image construction}
When the stego-data is transferred over a communication media, the intruder can access this data from the public channel and can try to construct the stego-image. If the intruder obtains a high visual quality image, then the goal of steganography is fulfilled. This is because he/ she will not be able to judge whether some data is hidden in the image or not. {\color{myblue}Therefore, in this subsection, we give the steps to construct the stego-image from the stego-data, which is equivalent to Section \ref{CS Reconstruction}. We refer this process as construction rather than reconstruction.}

\begin{enumerate}
	\item Obtain the approximate sparse vector ${s}'$ from the stego-data and measurement matrices $\Phi_{u}$ and $\Phi_{v}$ as (recall \eqref{eq: propose measurements})
\begin{linenomath*}
\begin{equation}
	\begin{aligned}
& {s}'_{i,u}=\Phi_{u}^{-1} y_{i,u}, \: and	\\ 
& {s}'_{i,v}= ADMM\_LASSO\left( z_{i,v}, \Phi_{v} \right). 
	\end{aligned}
\end{equation}
\end{linenomath*}
{\color{mygreen}Here, as discussed in Section \ref{CS Reconstruction}, we use ADMM and LASSO to construct ${s}'_{i,v}$.}  The sparse vector ${s}'$ is obtained by concatenating ${s}'_{i,u}$ and ${s}'_{i,v}$. Here, the size of ${s}'_{i,u}$, ${s}'_{i,v}$, and ${s}'$ is the same as that of $s_{i,u}$, $s_{i,v}$, and $s$, respectively.

	\item Convert each vector ${s}'_{i}$ into a block of size $B\times B$.  
	\item {\color{myblue}Apply two-dimensional Inverse DCT (IDCT) to each of these blocks to generate blocks ${x}'_{i}$ of image.} That is, recall \eqref{eq: propose sparsification},
\begin{linenomath*}
	\begin{equation}
	{x}'_{i}=IDCT\left( {s}'_{i}\right).
	\end{equation}
\end{linenomath*}
	\item Construct the stego-image of size $r1\times r2$ by arranging all these blocks ${x}'_{i}$. 
\end{enumerate}

\begin{figure}
	\centering
	\includegraphics[width=.45\textwidth]{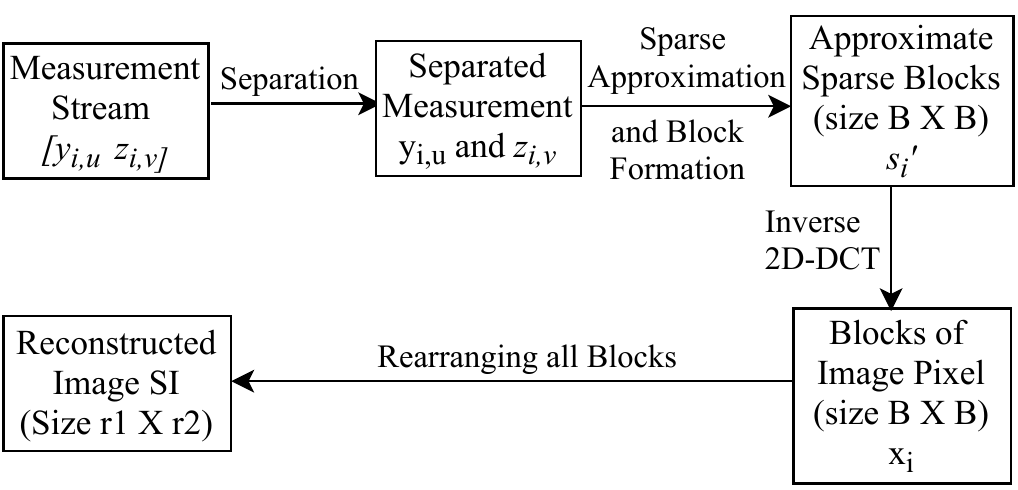}
	\caption{Stego-Image Construction}
	\label{Figure:Stego image construction}
\end{figure}

The block representation of these steps is given in Fig. \ref{Figure:Stego image construction}. We show in the Experimental Results section that image obtained from this stego-data preserves the quality of the original image. 

As earlier, we term our proposed steganography scheme as Compressed-Sensing-Image-Steganography (CSIS) because we use compressed sensing to enhance the embedding capacity of the image steganography scheme. 

\section{Experimental Results}\label{experimental results}
Experiments are carried out in MATLAB on a machine with an Intel Core i3 processor @2.30 GHz and 4GB RAM. {\color{myviolet}We use a set of standard grayscale images to test our CSIS. Sample test images are shown in Fig. \ref{Figure:test images} and Fig. \ref{Figure:test images Cont...}. These images have the varying texture property and are taken from the miscellaneous category of USC-SIPI image database \cite{SIPI} and two other public domain databases \cite{ImageDatabase1, ImageDatabase2}. 

The miscellaneous category of USC-SIPI database consists of 24 grayscale images. Some images, such as Lena, and Tiffany are no longer available in this database. These images have played a significant role in image processing, and literature. Thus, we use other public-domain test images databases \cite{ImageDatabase1, ImageDatabase2} for them. A total of seven such images are chosen. Hence, we have a total of 31 grayscale images.} Our CSIS is also applicable to color images, and we pick one of them from USC-SIPI database. 

{\color{myviolet}In this manuscript, we report average values of all the 31 images with detailed results for 10 images due to space limitations. This is further justified by the fact that the image processing literature has used these 10 images or a subset of them.


}

The size of each of test images is $512\times 512$, i.e. $r1\times r2$. We take blocks of size $8\times 8$, i.e. $B\times B$. As earlier, the size of measurement matrix $\Phi_u$ is $p_1\times p_1$. Recall from Section \ref{sec:data embedding}, $p_1$ is the number of coefficients with large values/ low frequency in the input sparse vector. For commonly used images, this value is between $10$ and $14$ \cite{Chang, NC}. Since the measurement matrix cannot be different for every input matrix, we do experiments with three different values of $p_1$ ($10$, $12$ and $14$) to find the optimal one here. Again from Section \ref{sec:data embedding}, the size of measurement matrix $\Phi_v$ is $(|m|-p_1)\times p_2$. We take $|m|$ from the following range \cite{Chang, NC}: $\{32, 35, 36, 37, 39, 40, 42, 47\}$, and as before, $p_2=B\times B-p_1$ (i.e. $p_2 = 64 - p_1$). For secret data, we use randomly generated data, which is sequence of $0$ and $1$ bits.

\textit{First}, we check the embedding capacity of our proposed scheme. \textit{Second}, we do the similarity analysis between the cover images and the constructed stego-images by assessing . \textit{Third}, in the remainder of this section, we do security analysis, perform five comparisons with existing steganography schemes, and also experiment with a color image.
 
\begin{figure*}[b]
	\centering
	\begin{subfigure}[b]{0.19\textwidth}
		\centering
		\includegraphics[width=2.5cm,height=3cm,keepaspectratio]{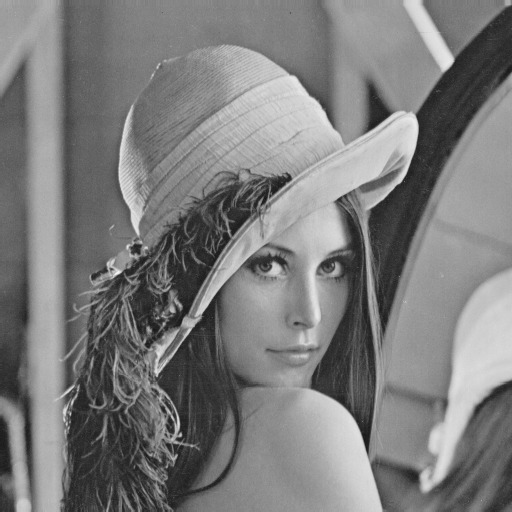}
		\caption{Lena}
		\label{fig:lena}
	\end{subfigure}
	\hfill
	\begin{subfigure}[b]{0.19\textwidth}
		\centering
		\includegraphics[width=2.5cm,height=3cm,keepaspectratio]{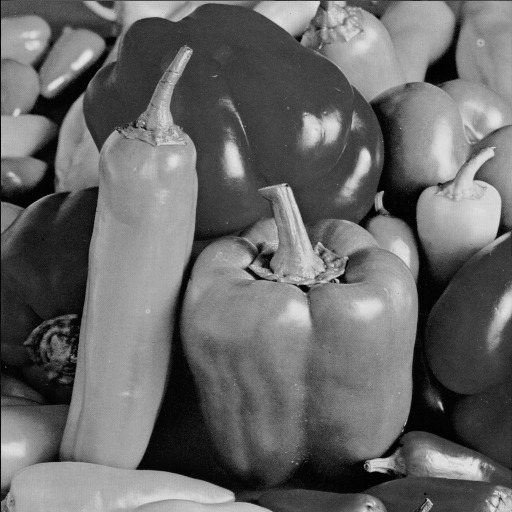}
		\caption{Peppers}
		\label{fig:peppers}
	\end{subfigure}
	\hfill
	\begin{subfigure}[b]{0.19\textwidth}
		\centering
		\includegraphics[width=2.5cm,height=3cm,keepaspectratio]{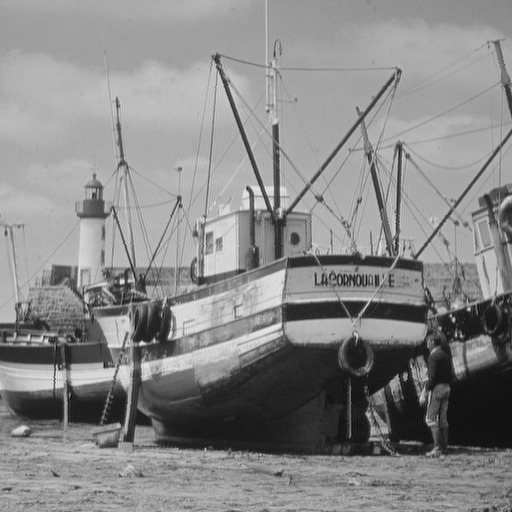}
		\caption{Boat}
		\label{fig:boat}
	\end{subfigure}
	\hfill
	\begin{subfigure}[b]{0.19\textwidth}
		\centering
		\includegraphics[width=2.5cm,height=3cm,keepaspectratio]{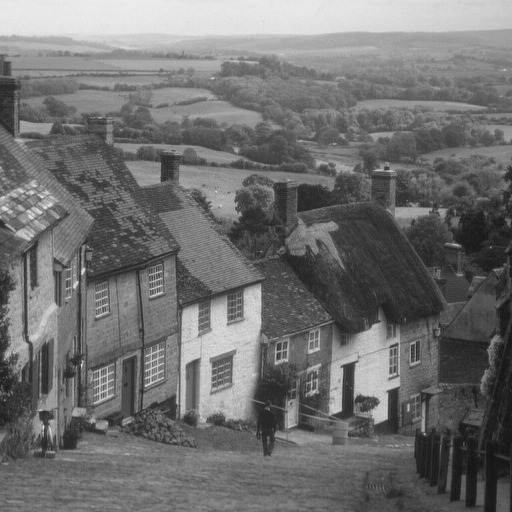}
		\caption{Goldhill}
		\label{fig:goldhill}
	\end{subfigure}
	\hfill
	\begin{subfigure}[b]{0.19\textwidth}
		\centering
		\includegraphics[width=2.5cm,height=3cm,keepaspectratio]{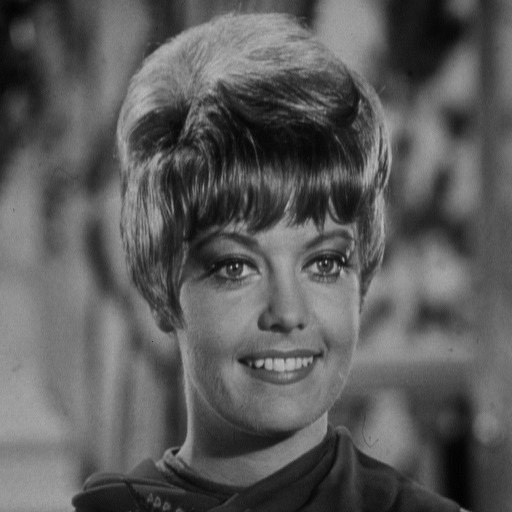}
		\caption{Zelda}
		\label{fig:zelda}
	\end{subfigure}
	\caption{Test images used in our experiments}
	\label{Figure:test images}
   \end{figure*}

\begin{figure*}[b]
	\centering
		\begin{subfigure}[b]{0.19\textwidth}
		\centering
		\includegraphics[width=2.5cm,height=3cm,keepaspectratio]{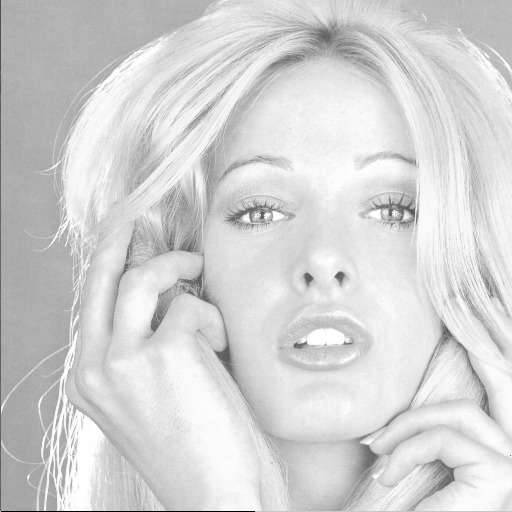}
		\caption{Tiffany}
		\label{fig:tiffany}
	\end{subfigure}
	\hfill
	\begin{subfigure}[b]{0.19\textwidth}
		\centering
		\includegraphics[width=2.5cm,height=3cm,keepaspectratio]{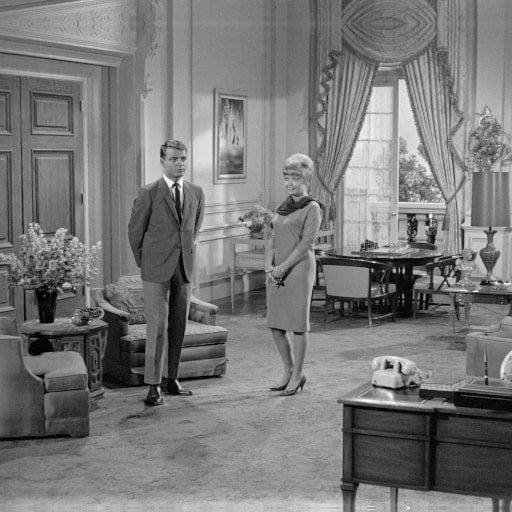}
		\caption{Living room}
		\label{fig:livingroom}
	\end{subfigure}
	\hfill
		\begin{subfigure}[b]{0.19\textwidth}
		\centering
		\includegraphics[width=2.5cm,height=3cm,keepaspectratio]{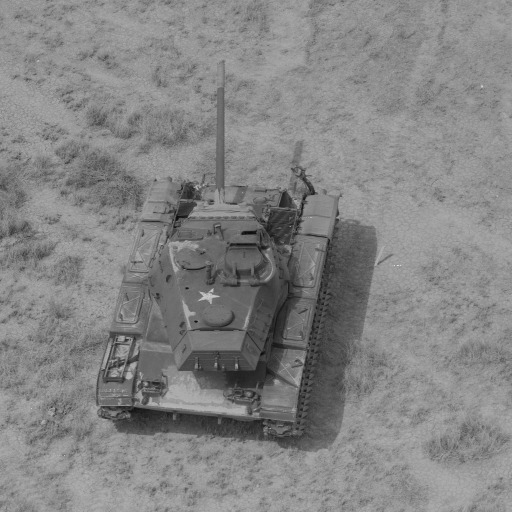}
		\caption{Tank}
		\label{fig:tank}
	\end{subfigure}
	\hfill
	\begin{subfigure}[b]{0.19\textwidth}
		\centering
		\includegraphics[width=2.5cm,height=3cm,keepaspectratio]{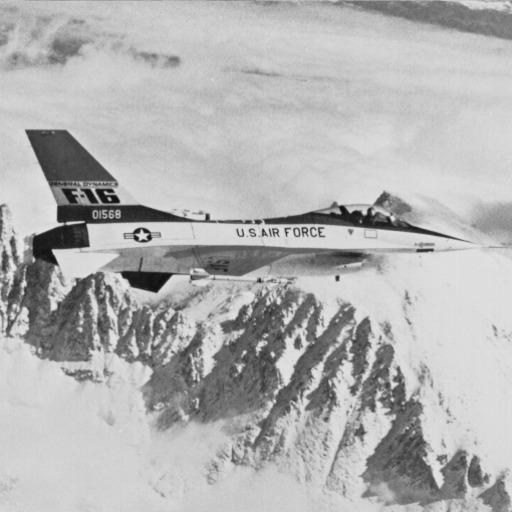}
		\caption{Airplane}
		\label{fig:jetplane}
	\end{subfigure}
	\hfill
	\begin{subfigure}[b]{0.19\textwidth}
		\centering
		\includegraphics[width=2.5cm,height=3cm,keepaspectratio]{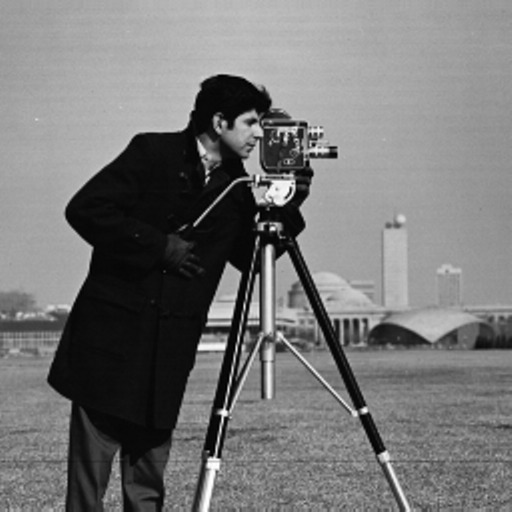}
		\caption{Camera man}
		\label{fig:cameraman}
	\end{subfigure}
	\caption{Continued from Fig. \ref{Figure:test images}; test images used in our experiments}
	\label{Figure:test images Cont...}
\end{figure*}

\subsection{Embedding Capacity Analysis}\label{embedding capacity analysis}
Embedding capacity is defined as the maximum number of bits embedded in the cover media, which is the image here. The embedding capacity of our proposed steganography scheme depends on the sampling rate (SR), which is given as
\begin{linenomath*}
\begin{alignat}{2}
& SR=\frac{\text{Total Linear Measurements}}{\text{Total Pixels in Cover Image}} .
\end{alignat}
\end{linenomath*}

We have $r_1 \times r_2$ total pixels in the cover image and $|m|$ linear measurements for each block with $\frac{r1\times r2}{B\times B}$ number of blocks. Therefore, our sampling rate is
\begin{linenomath*}
\begin{alignat}{2}
& SR=\left (  \frac{|m|}{r1\times r2}\right )\times \left (  \frac{r1\times r2}{B\times B} \right )=\frac{\left | m \right |}{B\times B}.
\end{alignat}
\end{linenomath*}

From this definition, it is evident that embedding capacity mainly depends upon $|m|$, however, the compressed image quality depends upon both $p_1$ and $|m|$. Therefore, to maintain the quality of stego-image while enhancing embedding capacity, the combination of these parameters is critical. 

{\color{myviolet}For different combinations of $p_1$ and $|m|$, in Table \ref{Table:Embedding capacity for different parameter}, we give the embedding capacity in bits of our proposed CSIS for the 10 test images of Fig. \ref{Figure:test images} and Fig. \ref{Figure:test images Cont...} and the average capacity for all the 31 images.} We analyze the data of this table by comparing $p_1$ and $|m|-p_1$ instead of $p_1$ and $|m|$ because the former set directly maps to the number of ordinary samples and compressed sensing samples, respectively. {\color{myblue}When $p_1$ is constant, and $|m|-p_1$ is increased, the number of compressed sensing samples increases, where the secret data bits are embedded, leading to increased capacity.} For example, consider columns $2$ and $3$ of Table \ref{Table:Embedding capacity for different parameter}, we can observe that the embedding capacity increases when $p_1$ is constant, i.e. $10$ and $|m|-p_1$ is increased from $22$ to $25$. When $|m|-p_1$ is constant and $p_1$ is increased, the number of compressed sensing samples decrease leading to decreased embedding capacity. For example, consider columns $3$ and $4$, we observe that embedding capacity decreases when $|m|-p_1$ is constant, i.e. $25$ and $p_1$ is increased from $10$ to $12$. 
\begin{table*}[h]
\fwprocesstable{Embedding capacity (in bits) obtain by proposed CSIS for different parameters and for different test images \label{Table:Embedding capacity for different parameter}}
{\begin{tabular*}{\textwidth}{@{\extracolsep{\fill}}lcccccccc}\toprule
\multirow{1}{*}{\begin{tabular}[c]{@{}l@{}}Test image\end{tabular}} & \multicolumn{8}{c}{Parameters}\\  
         & \begin{tabular}[c]{@{}c@{}}$p_1=10$\\ $|m|=32$\end{tabular} & \begin{tabular}[c]{@{}c@{}}$p_1=10$\\ $|m|=35$\end{tabular} & \begin{tabular}[c]{@{}c@{}}$p_1=12$\\ $|m|=37$\end{tabular} & \begin{tabular}[c]{@{}c@{}}$p_1=12$\\ $|m|=40$\end{tabular} & \begin{tabular}[c]{@{}c@{}}$p_1=12$\\ $|m|=42$\end{tabular} & \begin{tabular}[c]{@{}c@{}}$p_1=12$\\ $|m|=47$\end{tabular} & \begin{tabular}[c]{@{}c@{}}$p_1=14$\\ $|m|=36$\end{tabular} & \begin{tabular}[c]{@{}c@{}}$p_1=14$\\ $|m|=39$\end{tabular} \\
\midrule
Lena       & 171087   & 194519   & 194265   & 217491   & 232924   & 272130   & 170361 & 193679   \\ 
Peppers    & 173091   & 196725   & 196357   & 219641   & 235265   & 274890   & 172304 & 196193   \\ 
Boat       & 171563   & 194819   & 194559   & 217665   & 233430   & 272162   & 170738 & 194167   \\ 
Goldhill   & 174359   & 198019   & 197477   & 221155   & 236888   & 276297   & 173674 & 197031   \\ 
Zelda      & 170447   & 193811   & 193635   & 216639   & 232441   & 270830   & 170080 & 192951   \\ 
Tiffany    & 170457   & 193717   & 193291   & 216419   & 231924   & 270386   & 169747 & 192739   \\ 
Living room & 174534   & 198336   & 198216   & 222186   & 238076   & 277904   & 174402 & 198336   \\ 
Tank       & 174961   & 198933   & 198395   & 222165   & 238276   & 277972   & 174564 & 198223   \\ 
Airplane   & 167255   & 189865   & 189195   & 212003   & 227341   & 265207   & 165822 & 188313   \\ 
Camera man  & 161201   & 183181   & 180375   & 202601   & 215917   & 251596   & 157618 & 177801  \\  \midrule
Avg. of 10 images   & 170895 & 194192 & 193576 & 216796 & 232248 & 270937 & 169931 & 192943 \\
Avg. of 31 images & 152786 &	176645	& 174678 &	198080 &	214135 &	251989 &	150023 &	173564 \\
\botrule
\end{tabular*}}{}
\end{table*}

\subsection{Stego-image Quality Assessment}\label{similarity analysis}
In general, when the embedding capacity increases, the visual quality of stego-image degrades. Hence, with increased embedding capacity, preserving the visual quality of stego-image is also essential. There is no universal metric to judge the quality of stego-image. However, we check the quality of stego-image by examining the similarity between cover images and their corresponding stego-images. {\color{myblue}

This check is done in two ways. Initially we perform a visual or subjective check. The subjective measure is a good way to assess the quality of stego-image, but it depends on many factors like viewing distance, the display device, the lighting condition, viewer\textquotesingle s vision ability, and viewer\textquotesingle s mood. Therefore, it is necessary to design mathematical models to assess the quality of stego-images, which we discuss next.}

\subsubsection{Subjective or Visual Measure}\label{Visual similarity}

\begin{figure}[h]
	\centering
	\begin{subfigure}[b]{0.20\textwidth}
		\centering
		\includegraphics[width=2.5cm,height=3cm,keepaspectratio]{Peppers.jpg}
		\caption{`Pepper' cover image}
		\label{fig:pepper cover image}
	\end{subfigure}
	\begin{subfigure}[b]{0.20\textwidth}
		\centering
		\includegraphics[width=3cm,height=2.5cm]{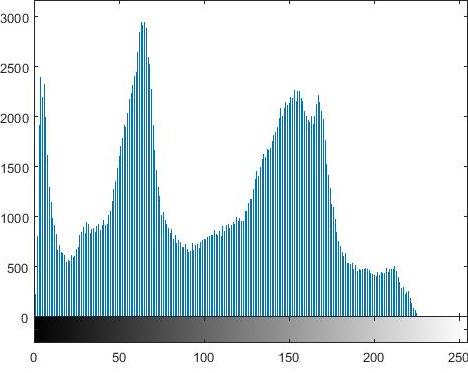}
		\caption{Cover image histogram}
		\label{fig:pepper cover image hist}
	\end{subfigure}
	\hfill
	\begin{subfigure}[b]{0.20\textwidth}
		\centering
		\includegraphics[width=2.5cm,height=3cm,keepaspectratio]{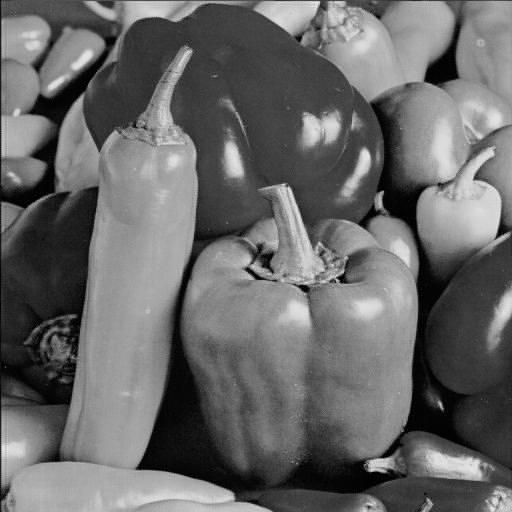}
		\caption{`Pepper' stego-image}
		\label{fig:pepper stego-image}
	\end{subfigure}
	\begin{subfigure}[b]{0.20\textwidth}
		\centering
		\includegraphics[width=3cm,height=2.5cm]{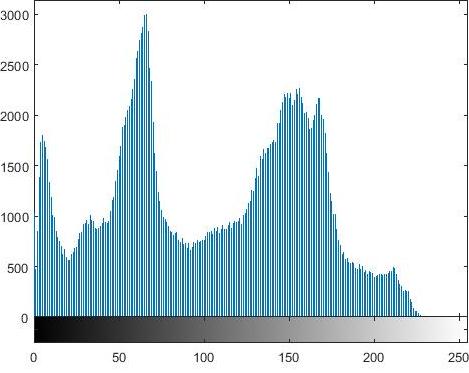}
		\caption{Stego-image histogram}
		\label{fig:pepper stego-image hist}
	\end{subfigure}
	\hfill
	\caption{`Pepper' cover image, its stego-image, and their corresponding histogram using parameter $p_1$=12 and $|m|$=37.}
	\label{Figure:visual analysis}
\end{figure}

Human observers are the final arbiter of image quality. Therefore, the subjective measure is a perfect way of assessing the quality of the images. Here, we construct stego-images corresponding to different test images used in our experiment for different combinations of $p_1$ and $|m|$. This result shows that the stego-images are almost similar to their corresponding cover images. The same is true for their corresponding histograms also. As an example, we present the visual comparison for `Pepper' cover image for one set of parameters; $p_1 = 12$ and $|m| = 37$. Fig. \ref{Figure:visual analysis} shows the (a) `Pepper' cover image (b) `Pepper' cover image histogram (c) `Pepper' stego-image (d) `Pepper' stego-image histogram. From these figures, we observe that the stego-image is almost similar to its corresponding cover image and their corresponding histograms are also very similar. 

\begin{figure}[h]
	\centering
	\begin{subfigure}[b]{0.20\textwidth}
		\centering
		\includegraphics[width=2.5cm,height=3cm,keepaspectratio]{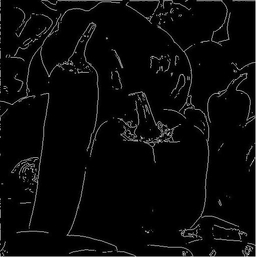}
		\caption{Cover image edge map}
		\label{fig:pepper cover edge map}
	\end{subfigure}
	\begin{subfigure}[b]{0.20\textwidth}
		\centering
		\includegraphics[width=2.5cm,height=3cm,keepaspectratio]{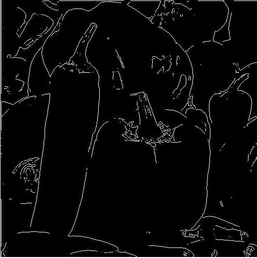}
		\caption{Stego-image edge map}
		\label{fig:pepper stego edge map}
	\end{subfigure}
	\caption{Edge maps of `Pepper' cover image and its stego-image using parameter $p_1$=12 and $|m|$=37.}
	\label{Fig:Edge map}
\end{figure}

We also construct the edge map diagrams for both the cover image and its corresponding stego-image for this same example. These edge maps are shown in Fig. \ref{fig:pepper cover edge map} and Fig. \ref{fig:pepper stego edge map}, respectively. We can see from these figures that both the edge maps are almost the same. Hence, the visual quality of the cover image and its corresponding stego-image is almost similar.
 
\subsubsection{Objective or Numerical Measures}\label{Numerical similarity}
These measures compare the cover images and their corresponding stego-images based on some numerical criteria that do not require extensive subjective studies. Hence, in recent times, these measures are more commonly used for image quality assessment. These include; Peak Signal-to-Noise Ratio (PSNR), mean Structural Similarity (SSIM) index, Normalized Cross-Correlation (NNC) coefficient, and entropy. We discuss all of them below.

\paragraph{\textbf{PSNR}}\label{PSNR}
We compute the \textit{PSNR} value to evaluate the imperceptibility of stego-images. That is,
\begin{equation}
PSNR=10\log_{10}\frac{R^{2}}{MSE}\: dB,
\end{equation}
where $MSE$ represents the mean square error between the cover image $I$ and the stego-image $SI$, $R$ is the maximum intensity of pixel, which is $255$ for grayscale images, and dB refer to decibel. The $MSE$ is calculated as
\begin{equation}
MSE=\frac{\sum_{i=1}^{r1}\sum_{j=1}^{r2}\left ( I\left ( i,\, j \right )-SI\left ( i,\, j \right ) \right )^2}{r1\times r2},
\end{equation}
where $r1$ and $r2$ represent the row and column numbers of the digital image, respectively, and $I(i,j)$ and  $SI(i,j)$ represent the pixel value of the cover image and the constructed stego-image, respectively. 

A higher PSNR value indicates the higher imperceptibility of the stego-image. In general, a value higher than 30 dB is considered to be good since human eyes can hardly distinguish the distortion in the stego-image \cite{Liu, Zhang2013}. The \textit{PSNR} values of the stego-images corresponding to 10 test images of Fig. \ref{Figure:test images} and \ref{Figure:test images Cont...}, and average for all 31 images for different combination of $p_1$ and $|m|$ are given in Table \ref{Table:PSNR for different parameter}. From this table, we can easily observe that this value is higher than 30 dB for all combinations of parameters and for all images. 

\begin{table*}[!h]
\fwprocesstable{Value of PSNR (in dB) obtain by proposed CSIS for different parameters and for different test images\label{Table:PSNR for different parameter}}
{\begin{tabular*}{\textwidth}{@{\extracolsep{\fill}}lcccccccc}\toprule
\multirow{1}{*}{\begin{tabular}[c]{@{}l@{}}Test image\end{tabular}} & \multicolumn{8}{c}{Parameters}\\ 
         & \begin{tabular}[c]{@{}c@{}}$p_1=10$\\ $|m|=32$\end{tabular} & \begin{tabular}[c]{@{}c@{}}$p_1=10$\\ $|m|=35$\end{tabular} & \begin{tabular}[c]{@{}c@{}}$p_1=12$\\ $|m|=37$\end{tabular} & \begin{tabular}[c]{@{}c@{}}$p_1=12$\\ $|m|=40$\end{tabular} & \begin{tabular}[c]{@{}c@{}}$p_1=12$\\ $|m|=42$\end{tabular} & \begin{tabular}[c]{@{}c@{}}$p_1=12$\\ $|m|=47$\end{tabular} & \begin{tabular}[c]{@{}c@{}}$p_1=14$\\ $|m|=36$\end{tabular} & \begin{tabular}[c]{@{}c@{}}$p_1=14$\\ $|m|=39$\end{tabular} \\
\midrule
Lena       & 34.34  & 35.11  & 35.62  & 36.15  & 36.71  & 37.31  & 36.33  & 36.91  \\ 
Peppers    & 34.05  & 34.35  & 35.23  & 35.76  & 36.21  & 36.98  & 35.44  & 35.81  \\ 
Boat       & 32.67  & 33.07  & 33.84  & 34.25  & 34.72  & 36.84  & 34.37  & 34.81  \\ 
Goldhill   & 32.69  & 33.61  & 34.06  & 34.54  & 35.12  & 35.32  & 34.33  & 34.93  \\ 
Zelda      & 39.31  & 39.61  & 40.10   & 41.32  & 40.02  & 42.67  & 40.73  & 42.46  \\ 
Tiffany    & 33.64  & 33.96  & 34.69  & 35.88  & 36.49  & 37.23  & 35.73  & 36.37  \\ 
Living room & 30.94  & 31.29  & 32.07  & 32.98  & 33.31  & 33.78  & 33.48  & 33.64  \\ 
Tank       & 34.27  & 34.32  & 35.13  & 35.62  & 35.98  & 36.98  & 35.36  & 35.87  \\ 
Airplane   & 32.89  & 34.15  & 34.88  & 35.78  & 36.39  & 37.91  & 34.43  & 35.42  \\ 
Camera man  & 35.71  & 36.89  & 37.52  & 38.86  & 39.38  & 40.86  & 40.04  & 40.65  \\ 
\midrule
Avg. of 10 images    & 34.051 & 34.636 & 35.314 & 36.114 & 36.433 & 37.588 & 36.024 & 36.687 \\
Avg. of 31 images &	34.245	& 34.883 &	35.593 &	36.379 &	36.668	& 37.921 &	36.282 &	36.901 \\
\botrule
\end{tabular*}}{}
\end{table*}

\paragraph{\textbf{Means SSIM Index}}\label{SSIM}

It is an image quality assessment metric used to measure the structural similarity between two images \cite{SSIM}. This measure is based on the assumption that the human visual system (HVS) is more adapted to the image\textquotesingle s structural information. The mean SSIM (MSSIM) index is given as
\begin{align}
& SSIM(x,y)=\frac{(2\mu _{x}\mu _{y}+C_{1})(2\sigma _{xy}+C_{2})}{(\mu _{x}^{2}+\mu _{y}^{2}+C_{1})(\sigma _{x}^{2}+\sigma _{y}^{2}+C_{2})}, \\
& MSSIM(I,SI)=\frac{1}{M}\sum_{j=1}^{M}SSIM(i_{j},si_{j}),
\end{align} 
{\color{myblue}where $SSIM(x,y)$ calculates the SSIM index for vectors $x$ and $y$, and $MSSIM$ $(I,SI)$ calculates the mean SSIM between cover image $I$ and stego-image $SI$, i.e. for the overall image quality}. Here, $\mu_{x}$ is the weighted mean of $x$, $\mu_{y}$ is the weighted mean of $y$, $\sigma_{x}$ is the weighted standard deviation of $x$, $\sigma_{y}$ is the weighted standard deviation of $y$, $\sigma_{xy}$ is the weighted covariance between $x$ and $y$, $C_{1}$ \& $C_{2}$ are {\color{myrosewood}arbitrary constants}, $i_{j}$ \& $si_{j}$ are the content of the cover image and stego-image, respectively, at the $j^{th}$ local window, and $M$ is the number of local windows. We took the values of all these parameters according to \cite{SSIM}. 
The value of the mean SSIM index lies between $0$ and $1$, where the value $0$ indicates that there is no similarity between the two images, and the value $1$ indicates that the images are exactly similar. 

The mean SSIM index values between the stego-images and their corresponding cover images for different combination of $p_1$ and $|m|$ are given in Table \ref{Table:SSIM for different parameter}. As earlier, 10 images from \ref{Figure:test images} and \ref{Figure:test images Cont...} are extensively analyze and average of 31 images is reported. From this table, we observe that all these values are close to $1$, which represents that the stego-images are very much similar in structure to their corresponding cover images.

\begin{table*}[!h]
\fwprocesstable{Value of Mean SSIM index obtain by proposed CSIS for different parameters and for different test images\label{Table:SSIM for different parameter}}
{\begin{tabular*}{\textwidth}{@{\extracolsep{\fill}}lcccccccc}\toprule
\multirow{1}{*}{\begin{tabular}[c]{@{}l@{}}Test image\end{tabular}} & \multicolumn{8}{c}{Parameter} \\  
        & \begin{tabular}[c]{@{}c@{}}$p_1=10$\\ $|m|=32$\end{tabular} & \begin{tabular}[c]{@{}c@{}}$p_1=10$\\ $|m|=35$\end{tabular} & \begin{tabular}[c]{@{}c@{}}$p_1=12$\\ $|m|=37$\end{tabular} & \begin{tabular}[c]{@{}c@{}}$p_1=12$\\ $|m|=40$\end{tabular} & \begin{tabular}[c]{@{}c@{}}$p_1=12$\\ $|m|=42$\end{tabular} & \begin{tabular}[c]{@{}c@{}}$p_1=12$\\ $|m|=47$\end{tabular} & \begin{tabular}[c]{@{}c@{}}$p_1=14$\\ $|m|=36$\end{tabular} & \begin{tabular}[c]{@{}c@{}}$p_1=14$\\ $|m|=39$\end{tabular} \\
\midrule
Lena  & 0.9308 & 0.9394   & 0.9475            & 0.9518            & 0.9562             & 0.9672            & 0.9512            & 0.9558            \\ 
Peppers     & 0.9203 & 0.9225            & 0.9291  & 0.9463     & 0.9424            & 0.9547            & 0.9333           & 0.9383            \\ 
Boat        & 0.9211 & 0.9356            & 0.9444            & 0.9517            & 0.9575            & 0.9663            & 0.9484             & 0.9545            \\ 
Goldhill    & 0.9011 & 0.9122             & 0.9236            & 0.9343            & 0.9421             & 0.9532            & 0.9227            & 0.9359            \\ 
Zelda       & 0.9512 & 0.9563            & 0.9613            & 0.9657            & 0.9694            & 0.9768            & 0.9628            & 0.9678            \\
Tiffany     & 0.9239 & 0.9315            & 0.9357            & 0.9434            & 0.9501            & 0.9596            & 0.9369            & 0.9437            \\ 
Living room & 0.9012    & 0.9092             & 0.9211             & 0.9332             & 0.9384             & 0.9460             & 0.9341             & 0.9382             \\ 
Tank        & 0.8835 & 0.8857            & 0.9006            & 0.9094            & 0.9197            & 0.9388            & 0.9086            & 0.9131            \\ 
Airplane    & 0.9463 & 0.9525            & 0.9605            & 0.9646            & 0.9673            & 0.9755            & 0.9607            & 0.9692            \\ 
Camera  man & 0.9677 & 0.9752            & 0.9839            & 0.9864            & 0.9871            & 0.9907            & 0.9838            & 0.9868            \\ 
\midrule
Avg. of 10 images     & 0.9198 & 0.9272 & 0.9360 & 0.9445 & 0.9492 & 0.9598 & 0.9398 & 0.9463 \\
Avg. of 31 Images &	0.9206	& 0.9276 &	0.9365 &	0.9449 &	0.9498 &	0.9601 &	0.9412 &	0.9471 \\
\botrule
\end{tabular*}}{}
\end{table*}

\paragraph{\textbf{NCC Coefficient}}\label{NNC}

Normalized correlation (NC) metric measures the degree of similarity between two images, and when the two images are independent, this correlation is called normalized cross-correlation (NCC) \cite{NC}. The NCC coefficient is given as
\begin{align}
& NCC=\frac{\sum_{i=1}^{r1}\sum_{j=1}^{r2}I(i,j)SI(i,j)}{\sum_{i=1}^{r1}\sum_{j=1}^{r2}I^{2}(i,j)},
\end{align} 
where $r1$ and $r2$ represent the row and column numbers of the digital image, respectively. $I(i,j)$ and  $SI(i,j)$ represent the pixel value of the cover image and the constructed stego-image, respectively. The value equal to 1 indicates that both the images are exactly similar. For our experiments, the values of NCC are given in Table \ref{Table:NC}. The set of images used are same as for PSNR and SSIM. We observe that all these values are close to 1, which means that the stego-images are almost identical to their corresponding cover images. 

\begin{table*}[!h]
\fwprocesstable{Value of normalized cross-correlation obtain by proposed CSIS for different parameters and for different test images\label{Table:NC}}
{\begin{tabular*}{\textwidth}{@{\extracolsep{\fill}}lcccccccc}\toprule
\multirow{1}{*}{\begin{tabular}[c]{@{}l@{}}Test image\end{tabular}} & \multicolumn{8}{c}{Parameter} \\ 
        & \begin{tabular}[c]{@{}c@{}}$p_1=10$\\ $|m|=32$\end{tabular} & \begin{tabular}[c]{@{}c@{}}$p_1=10$\\ $|m|=35$\end{tabular} & \begin{tabular}[c]{@{}c@{}}$p_1=12$\\ $|m|=37$\end{tabular} & \begin{tabular}[c]{@{}c@{}}$p_1=12$\\ $|m|=40$\end{tabular} & \begin{tabular}[c]{@{}c@{}}$p_1=12$\\ $|m|=42$\end{tabular} & \begin{tabular}[c]{@{}c@{}}$p_1=12$\\ $|m|=47$\end{tabular} & \begin{tabular}[c]{@{}c@{}}$p_1=14$\\ $|m|=36$\end{tabular} & \begin{tabular}[c]{@{}c@{}}$p_1=14$\\ $|m|=39$\end{tabular} \\
\midrule
Lena       & 0.9985  & 0.9988              & 0.9989         & 0.9991               & 0.9991              & 0.9992              & 0.9991              & 0.9993              \\ 
Peppers    & 0.9982               & 0.9983              & 0.9985              & 0.9987              & 0.9988              & 0.9989              & 0.9985              & 0.9988              \\ 
Boat       & 0.9979              & 0.9983              & 0.9985              & 0.9987              & 0.9988              & 0.9989              & 0.9987              & 0.9989              \\ 
Goldhill   & 0.9976              & 0.9981               & 0.9983              & 0.9986              & 0.9987              & 0.9988              & 0.9982              & 0.9986              \\ 
Zelda      & 0.9991              & 0.9993              & 0.9994              & 0.9995              & 0.9995              & 0.9997              & 0.9994              & 0.9995              \\ 
Tiffany    & 0.9992              & 0.9993              & 0.9994              & 0.9995              & 0.9995              & 0.9996              & 0.9994              & 0.9995              \\ 
Living room & 0.9962               & 0.9961               & 0.9962 & 0.9971 & 0.9972               & 0.9964               & 0.9970               & 0.9982               \\ 
Tank       & 0.9987              & 0.9988              & 0.9992               & 0.9991              & 0.9992              & 0.9993              & 0.9993               & 0.9991              \\ 
Airplane   & 0.9989              & 0.9991              & 0.9993              & 0.9994              & 0.9994              & 0.9995              & 0.9993              & 0.9995              \\ 
Cameraman  & 0.9989              & 0.9991              & 0.9994              & 0.9994              & 0.9994              & 0.9995              & 0.9995              & 0.9996              \\
\midrule
Avg. of 10 images    & 0.9983             & 0.9985              & 0.9989  & 0.9989             & 0.9989             & 0.9989             & 0.9988             & 0.9991             \\
Avg. of 31 images &	0.9983 &	0.9985	& 0.9990 &	0.9989	& 0.9989 &	0.9989 &	0.9990 &	0.9991 \\
\botrule
\end{tabular*}}{}
\end{table*}

\paragraph{\textbf{Entropy}}\label{Entropy}

In general, entropy is defined as the measure of average uncertainty of a random variable, which here is the average number of bits required to describe the random variable. In the context of an image, it is a statistical measure of randomness that can be used to characterize the texture of the image \cite{Gonzalez}. For a grayscale image, entropy is given as  
\begin{linenomath*}
\begin{align}
& Entropy=-\sum_{i=0}^{255}(p_{i}\log _{2}p_{i}), 
\end{align} 
\end{linenomath*}
where $p_{i}$ is the probability of value $i$ pixel of the image. Table \ref{Table:Entropy} gives the entropy values for the cover images and their corresponding stego-images for different combinations of $p_1$ and $|m|$. The set of images used are same as for PSNR, SSIM, and NCC. From this table, we observe that for all these combinations of $p_1$ and $|m|$, the entropy of the cover images and their corresponding stego-images are almost similar.

\begin{table*}[!h]
\fwprocesstable{Entropy comparison of cover images and their corresponding stego-images obtain by proposed CSIS using different parameters\label{Table:Entropy}}
{\begin{tabular*}{\textwidth}{@{\extracolsep{\fill}}lccccccccc}\toprule
	\multirow{2}{*}{\begin{tabular}[l]{@{}c@{}}Test \\ image\end{tabular}} & \multirow{2}{*}{\begin{tabular}[c]{@{}c@{}}Cover\\ image\end{tabular}} & \multicolumn{8}{c}{Stego-image using different parameters}  \\ 
         &       & \begin{tabular}[c]{@{}c@{}}$p_1=10$\\ $|m|=32$\end{tabular} & \begin{tabular}[c]{@{}c@{}}$p_1=10$\\ $|m|=35$\end{tabular} & \begin{tabular}[c]{@{}c@{}}$p_1=12$\\ $|m|=37$\end{tabular} & \begin{tabular}[c]{@{}c@{}}$p_1=12$\\ $|m|=40$\end{tabular} & \begin{tabular}[c]{@{}c@{}}$p_1=12$\\ $|m|=42$\end{tabular} & \begin{tabular}[c]{@{}c@{}}$p_1=12$\\ $|m|=47$\end{tabular} & \begin{tabular}[c]{@{}c@{}}$p_1=14$\\ $|m|=36$\end{tabular} & \begin{tabular}[c]{@{}c@{}}$p_1=14$\\ $|m|=39$\end{tabular} \\
\midrule
Lena       & 7.4456  & 7.4552  & 7.4581            & 7..4569          & 7.456   & 7.4545  & 7.4534  & 7.4551  & 7.4536  \\ 
Peppers    & 7.5715  & 7.5924  & 7.5924            & 7.5908           & 7.5911  & 7.5901  & 7.5889  & 7.5897  & 7.5898  \\ 
Boat       & 7.1238  & 7.1323  & 7.1339            & 7.1322            & 7.1334  & 7.1337  & 7.1331  & 7.1277  & 7.1304  \\ 
Goldhill   & 7.4778  & 7.4653  & 7.4686            & 7.4704           & 7.4723  & 7.4719  & 7.4731  & 7.469   & 7.4717  \\ 
Zelda      & 7.2668  & 7.2625  & 7.2635            & 7.2638           & 7.2643  & 7.2649  & 7.2652  & 7.2633  & 7.2642  \\ 
Tiffany    & 6.6015  & 6.6076  & 6.6063            & 6.6046           & 6.606   & 6.6074  & 6.607   & 6.6096  & 6.6076  \\ 
Living room & 7.2950   & 7.4200 & 7.4200         & 7.4253            & 7.4260   & 7.4261   & 7.4262   & 7.4267   & 7.4278  \\
Tank       & 5.4957  & 6.3614  & 6.3728            & 6.3771           & 6.3829  & 6.3846  & 6.3871  & 6.3709  & 6.3815  \\ 
Airplane   & 6.7025  & 6.773   & 6.7637            & 6.7535           & 6.7501  & 6.7468  & 6.7396  & 6.7614  & 6.7454  \\ 
Camera man  & 7.0482   & 7.0743   & 7.0763             & 7.0738           & 7.0703  & 7.0683  & 7.0664  & 7.0726  & 7.0661  \\ 
\midrule
Avg. of 10 images    & 7.0028 & 7.1144 & 7.0691            & 7.07678 & 7.1152 & 7.1148 & 7.1140 & 7.1145 & 7.1138 
             \\
Avg. of 31 images &	6.9985 &	6.6451 &	7.7132 &	6.7124 &	6.6476	& 6.6462 &	6.6447 &	6.6469 &	6.6448 \\
\botrule
\end{tabular*}}{}
\end{table*}

\subsection{Security Analysis}\label{Security Analysis}
Since the proposed CSIS is a transform domain based technique and it employs indirect embedding strategy, i.e. it does not follow the LSB flipping method,and hence, it is immune to statistical attacks \cite{Westfeld, PM1_steganography}. {\color{myblue}Also, CSIS does not lead to the shrinkage effect. That means, after embedding, the nonzero coefficients do not modify to zero value, and hence attacks against F5 \cite{Fridrich, PM1_steganography} are not considered.} 

Moreover, in CSIS, the measurement matrix $\Phi$ is considered as the secret-key, which is shared between the sender and the legitimate receiver. 
If the eavesdropper intercepts the stego-image by a randomly generated measurement matrix, he cannot not enter the embedding domain without the original secret-key. Hence, we achieve increased security in our proposed system. To justify this, we extract the secret data in two ways, i.e. by using the correct measurement matrix and by using a measurement matrix that is very close to the original one, and obtain the BER (discussed in Section \ref{sec:data extraction}) between the original secret data the extracted one. 

\begin{figure}[]
	\centering
		\includegraphics[width=0.45\textwidth]{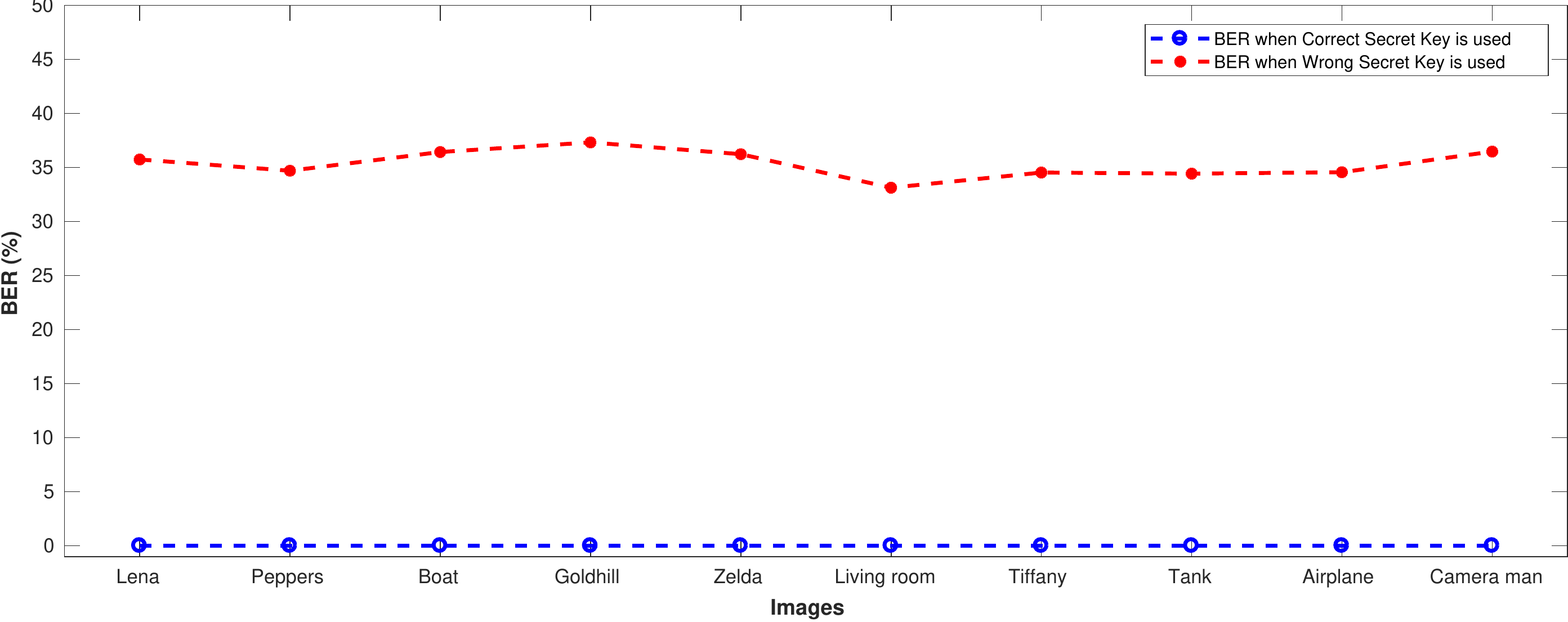}
		\caption{BER with the correct and with a wrong secret-key (i.e. measurement matrix)}
		\label{Fig:BER with a wrong secret-key (i.e. measurement matrix)}
\end{figure}

In Fig. \ref{Fig:BER with a wrong secret-key (i.e. measurement matrix)}, we present this BER for earlier discussed 10 cover images, and for the parameter $p_1$=12 and $|m|$=37. In this figure, we see that for the correct secret-key, the BER is 0, and for a tiny difference in the measurement matrix, i.e. wrong secret-key, the BER is very high, which is 35\% to 40\%. That is, even a small change in the secret-key will lead to an extreme shift in accuracy between the original secret data and the extracted one. 

\begin{figure}[]
	\centering 
	\begin{subfigure}[b]{0.45\textwidth}
		\centering
		\includegraphics[width=\textwidth]{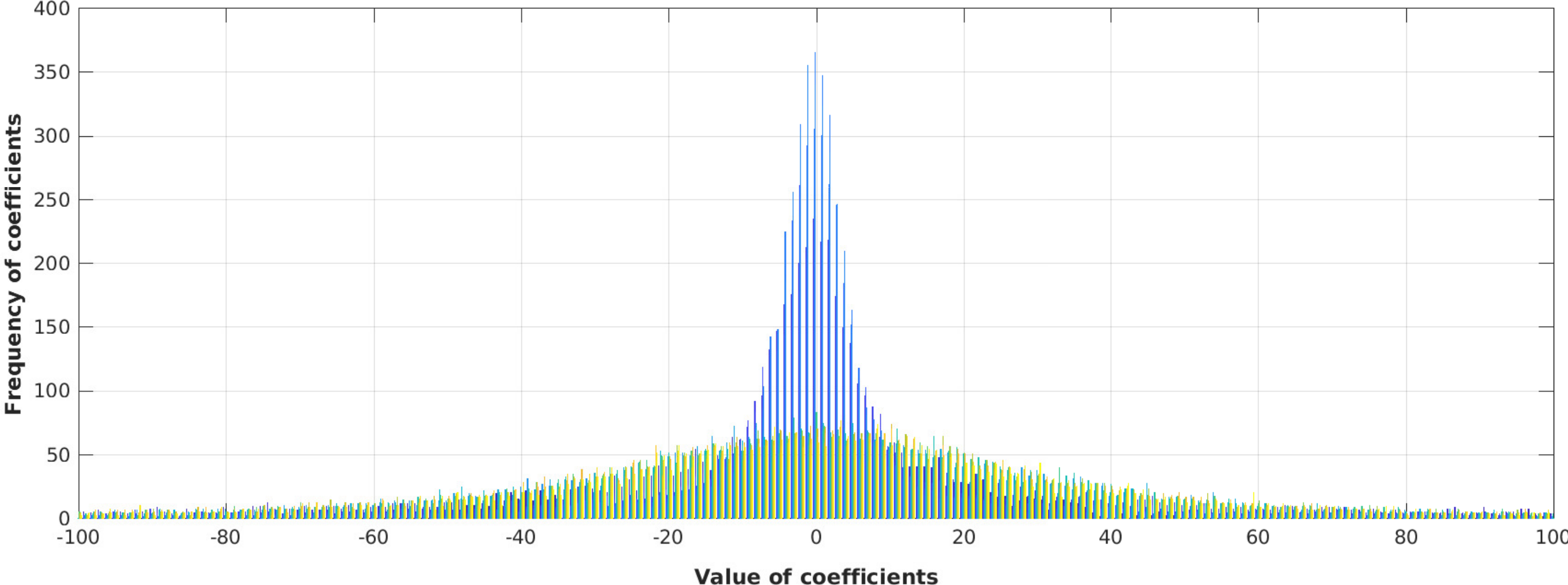}
		\caption{Original measurements}
		\label{Fig:Distribution of original transform coefficients}
	\end{subfigure}
	\hfill
		\begin{subfigure}[b]{0.45\textwidth}
		\centering
		\includegraphics[width=\textwidth]{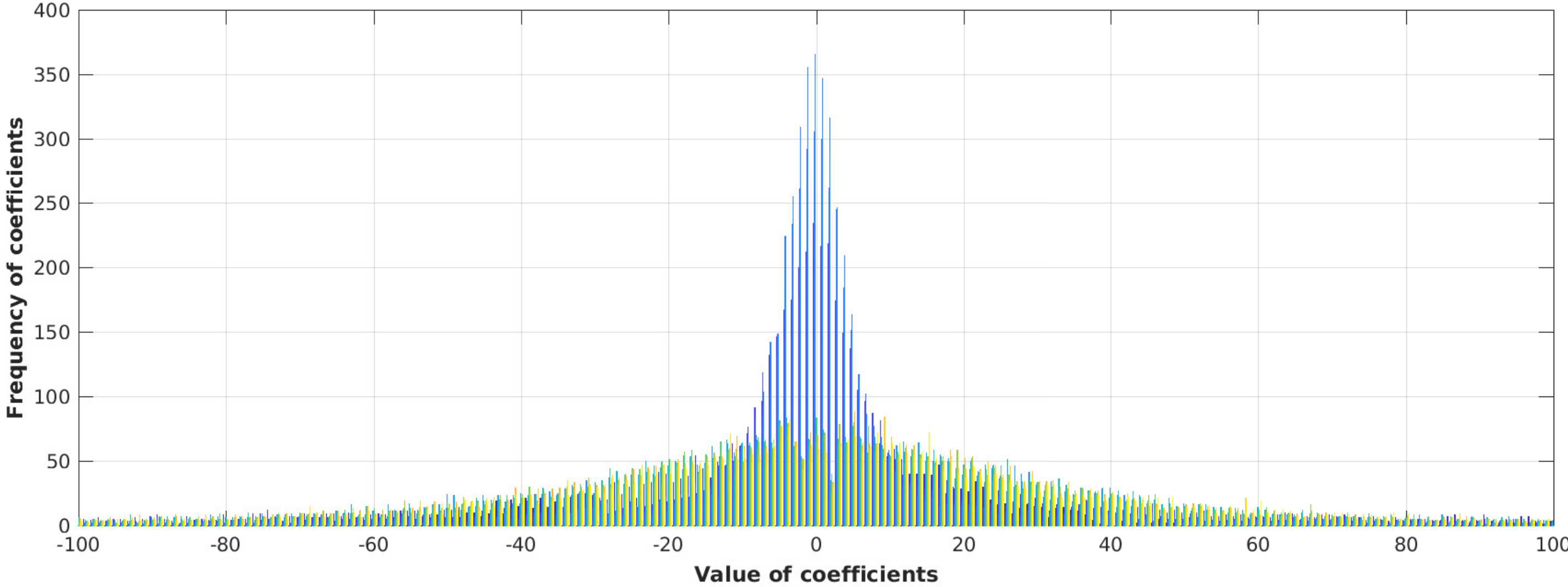}
		\caption{Modified measurements}
		\label{Fig:Distribution of modified transform coefficients}
	\end{subfigure}
	\caption{Distribution of measurements for `Peppers' image}
	\label{Fig:Hist in Transfrom domian}
\end{figure}

In addition to the above security analysis, we also measure the security by analyzing the distribution of the measurements and their corresponding modified measurements, i.e. after embedding the secret data. {\color{myblue}For `Pepper' image with parameter $p_1$=12 and $|m|$=37, this distribution of the original measurements and the modified measurements is shown in Fig. \ref{Fig:Distribution of original transform coefficients} and Fig. \ref{Fig:Distribution of modified transform coefficients}, respectively.} The green and blue colors are automatically added by Matlab and do not have any significance here. From these figures, we see that the distribution for both cases is almost the same. We also check these distributions for all the images and obtain the same results. We do not include these in this manuscript due to space limitations.

\begin{figure}[!h]
	\centering
		\includegraphics[width=0.45\textwidth]{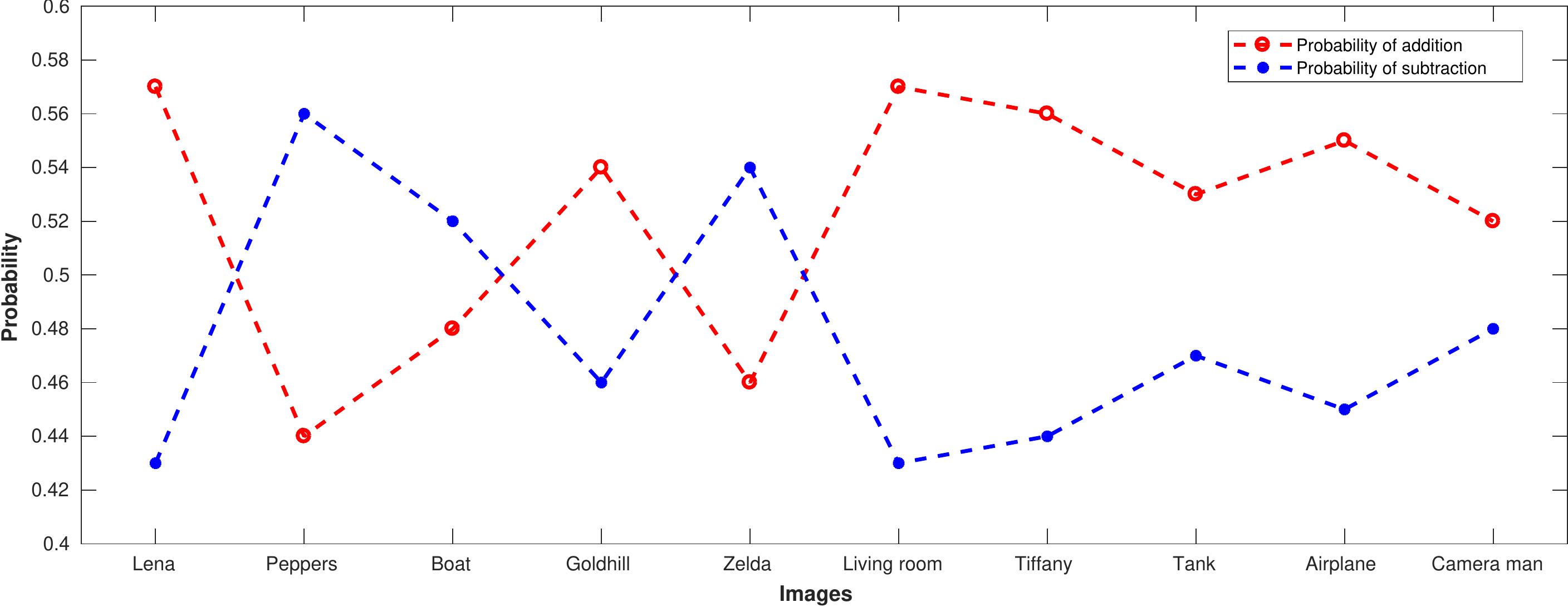}
		\caption{Probability of addition and subtraction operation}
		\label{Fig:Probability of addition and subtraction operation}
\end{figure}

The preservation of distribution of measurements in the earlier two histograms can also be justified by the probability of addition and subtraction operation decided by our algorithm. In Fig. \ref{Fig:Probability of addition and subtraction operation}, we plot this probability. From this figure, we see that the lines of probabilities of addition and subtraction operation oscillate around 0.5. Here, the minimum and maximum deviation to 0.5 are 0.02 and 0.07, respectively, i.e. for proposed CSIS, the probabilities of both the addition and the subtraction are nearly the same. The distribution of measurements and the probability of addition \& subtraction operation as discussed have justified that for our proposed CSIS, the likelihood of detecting data embedding by an eavesdropper is significantly low.
\subsection{Performance Comparison}\label{Performance Comparison}

In this subsection, we compare the performance of the proposed CSIS with the existing steganography schemes. This result is given in Table \ref{Table:Performance comparison with various other steganography schemes}. In this table, the first column represents the comparison metrics, and the remaining columns give the metric data for different steganography schemes. 
\begin{table*}[!h]
\fwprocesstable{Performance comparison between proposed CSIS and various other steganography schemes\label{Table:Performance comparison with various other steganography schemes}}
{\begin{tabular*}{\textwidth}{@{\extracolsep{\fill}}lccccccccc}\toprule
	\multirow{2}{*}{\textbf{Metrics}}  & \multicolumn{9}{c}{\textbf{Steganography Schemes}} \\ 
         & \textbf{CSIS} & {\color{myblue}Ref. \cite{Steg_2019_FGT}} & {\color{myblue}Ref. \cite{Jsteg}} & {\color{myblue}Ref. \cite{F5}} & {\color{myblue}Ref. \cite{Outguess}} & {\color{myblue}Ref. \cite{Liu}} & {\color{myblue}Ref. \cite{Pan}} & {\color{myblue}Ref. \cite{Steg_ExpertSystem}} & {\color{myblue}Ref. \cite{Steg_Rohit}}   \\
\midrule
\multicolumn{1}{l}{\begin{tabular}[l]{@{}l@{}}Capacity (in bits)\end{tabular}}                 & \multicolumn{1}{c}{\textbf{174678}} & \multicolumn{1}{c}{262144}  & \multicolumn{1}{c}{41267}               & \multicolumn{1}{c}{41451}         & \multicolumn{1}{c}{20644} & \multicolumn{1}{c}{55001} & \multicolumn{1}{c}{950}             & \multicolumn{1}{c}{57568}                            & \multicolumn{1}{c}{113960}                                             \\ 

\multicolumn{1}{l}{\begin{tabular}[l]{@{}l@{}}PSNR (in dB)\end{tabular}}                 & \multicolumn{1}{c}{\textbf{30.94 to 40.86}} & \multicolumn{1}{c}{36.51} & \multicolumn{1}{c}{34.39}               & \multicolumn{1}{c}{35.00}         & \multicolumn{1}{c}{34.54}  & \multicolumn{1}{c}{32.54}             & \multicolumn{1}{c}{35.52}             & \multicolumn{1}{c}{49.89}                            & \multicolumn{1}{c}{36.64}                                              \\ 

\multicolumn{1}{l}{\begin{tabular}[l]{@{}l@{}}Compression\\ Based\end{tabular}}            & \multicolumn{1}{c}{\textbf{Yes}} & \multicolumn{1}{c}{No} & \multicolumn{1}{c}{Yes}               & \multicolumn{1}{c}{Yes}         & \multicolumn{1}{c}{Yes}                     & \multicolumn{1}{c}{Yes}             & \multicolumn{1}{c}{Yes}             & \multicolumn{1}{c}{No}                            & \multicolumn{1}{c}{Yes}                                             \\ 

\multicolumn{1}{l}{\begin{tabular}[l]{@{}l@{}}Resistant to\\ Chi-square\end{tabular}}         & \multicolumn{1}{c}{\textbf{Yes}} & \multicolumn{1}{c}{No} & \multicolumn{1}{c}{No} & \multicolumn{1}{c}{Yes}  & \multicolumn{1}{c}{Yes} & \multicolumn{1}{c}{Yes}             & \multicolumn{1}{c}{Yes}             & \multicolumn{1}{c}{Yes}                            & \multicolumn{1}{c}{Yes}                                              \\ 

\multicolumn{1}{l}{\begin{tabular}[l]{@{}l@{}}Resistant to\\ Shrinkage\\ Effect\end{tabular}} & \multicolumn{1}{c}{\textbf{Yes}} & \multicolumn{1}{c}{NA} & \multicolumn{1}{c}{Yes}  & \multicolumn{1}{c}{No}         & \multicolumn{1}{c}{Yes} & \multicolumn{1}{c}{Yes}             & \multicolumn{1}{c}{Yes}    & \multicolumn{1}{c}{Yes}                            & \multicolumn{1}{c}{Yes}                                              \\ 

\multicolumn{1}{l}{\begin{tabular}[l]{@{}l@{}}Secret Key\end{tabular}} & \multicolumn{1}{c}{\textbf{Yes}}  & \multicolumn{1}{c}{No} & \multicolumn{1}{c}{No}    & \multicolumn{1}{c}{No}   & \multicolumn{1}{c}{No}                     & \multicolumn{1}{c}{No}             & \multicolumn{1}{c}{Yes}             & \multicolumn{1}{c}{No}                            & \multicolumn{1}{c}{No}                                             \\
\botrule
\end{tabular*}}{}
\end{table*}

In the first row of Table \ref{Table:Performance comparison with various other steganography schemes}, we compare the average embedding capacity over all the 31 images. We report these embedding capacity for the parameter $p_1 = 12$ \& $|m|=37$. In this table, we do not compare these results for all the images because the existing schemes\textquotesingle{} data are not available for all the images. {\color{myblue}From the first row of this table, we observe that on an average our steganography scheme has approximately $0.67$, $4.23$, $4.21$, $8.46$, $3.18$, $183.87$, $3.03$ and $1.53$ times embedding capacity as compared to references \cite{Steg_2019_FGT}, \cite{Jsteg}, \cite{F5}, \cite{Outguess}, \cite{Liu},  \cite{Pan}, \cite{Steg_ExpertSystem}, and \cite{Steg_Rohit}, respectively.} {\color{myviolet}Here, we can see that our proposed scheme has a higher embedding capacity compared to all schemes except the one, which is \cite{Steg_2019_FGT}. The reason for this is that this scheme is based on embedding secret data in the spatial domain. As discussed in the Introduction, spatial domain based embedding techniques have a higher embedding capacity, but they are prone to security issues. Also, these techniques are not based on compression, which is the main motivation of this manuscript. Further, as evident from Table \ref{Table:Embedding capacity for different parameter}, for a set of parameters $p_1=12$ and $|m|=47$, CSIS has 270937, and 251989 bits embedding capacity for the average of 10 and 31 images, respectively. Hence, for this set of parameters, CSIS has approximately the same embedding capacity as that of \cite{Steg_2019_FGT}.} 

In the second row of this table, for our scheme we report the range of PSNR values when considering all sets of parameters and again all 31 images. From the second row of this table, we observe that similar to existing steganography schemes, our CSIS also has PSNR values greater than 30 dB, which is considered good \cite{Liu,Zhang2013}.

The purpose of the proposed CSIS is to embed secret data in the compressed domain. Hence, in the third row of Table \ref{Table:Performance comparison with various other steganography schemes}, we check which schemes are based on compression and which are not. From this row, we observe that except \cite{Steg_2019_FGT, Steg_ExpertSystem},  our CSIS and all other schemes are based on compression. Finally, from the fourth row to the sixth row of Table \ref{Table:Performance comparison with various other steganography schemes}, we compare the security of these schemes by checking whether they are resistant to chi-square attack or not, resistant to shrinkage effect or not, and use any secret-key or not. We observe that only our proposed CSIS and \cite{Pan} schemes pass all the three security tests. Hence, we can conclude that out of all these schemes, only CSIS fulfills all the goals of steganography with higher embedding capacity.  



\subsection{Experiments on Color Image}\label{Experiment on Color Image}
All the above experiments were performed on the grayscale images. However, we also show the applicability of our proposed CSIS on a color image. For this we only use `Pepper' color image of resolution $512 \times 512$, and perform experiments for $p_1=12$ and $|m|=37$ as well as $p_1=14$ and $|m|=36$. 
\begin{table*}[!h]
\fwprocesstable{The performance analysis of our proposed scheme on color cover image ($512 \times 512$ Pepper color image) using different parameters.\label{Table:performance analysis of color image}}
{\begin{tabular*}{\textwidth}{@{\extracolsep{\fill}}lccccccccc}\toprule
	\multirow{2}{*}{\begin{tabular}[c]{@{}c@{}}Parameters\end{tabular}} & \multirow{2}{*}{\begin{tabular}[c]{@{}c@{}}Embedding\\ capacity\end{tabular}} & \multirow{2}{*}{PSNR}  & \multirow{2}{*}{Mean SSIM} & \multicolumn{4}{c}{\begin{tabular}[c]{@{}c@{}}Normalized cross-correlation (NCC)\\ (for different color component)\end{tabular}} & \multicolumn{2}{c}{Entropy}   \\ 
		&                       &                                                                              &                       & Red                    & Green                  & Blue                   & Average                & \begin{tabular}[c]{@{}c@{}}Cover\\ image\end{tabular} & \begin{tabular}[c]{@{}c@{}}Stego\\ image\end{tabular} \\ \midrule
		\begin{tabular}[c]{@{}c@{}}$p_1=12$, $|m|=37$\end{tabular}      & 503863      & 33.89        
 & 0.9913 & 0.9990         & 0.9981               & 0.9963                & 0.9978               & 7.669                                           & 7.723                                                \\ 
		\begin{tabular}[c]{@{}c@{}}$p_1=14$, $|m|=36$\end{tabular}     & 573657  & 33.71                 
        & 0.9843                & 0.9989     & 0.9979                & 0.9960     & 0.9976    & 7.669                                               & 7.724                                                 \\ \midrule
		Average & 538760 & 33.80   & 0.9878 & 0.9990	& 0.9980 &	0.9962 & 0.9978 &	7.669 &	7.723 
             \\
\botrule
\end{tabular*}}{}
\end{table*}

\begin{figure}[!h]
	\centering
	\begin{subfigure}[b]{0.20\textwidth}
		\centering
		\includegraphics[width=2.5cm,height=3cm,keepaspectratio]{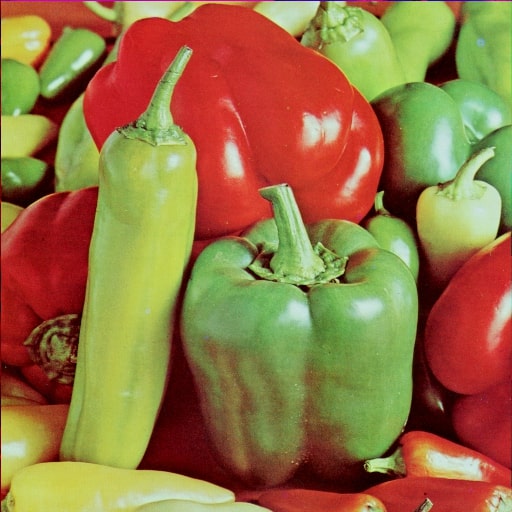}
		\caption{Color cover image}
		\label{fig:pepper color cover image}
	\end{subfigure}
	\begin{subfigure}[b]{0.20\textwidth}
		\centering
		\includegraphics[width=2.5cm,height=3cm,keepaspectratio]{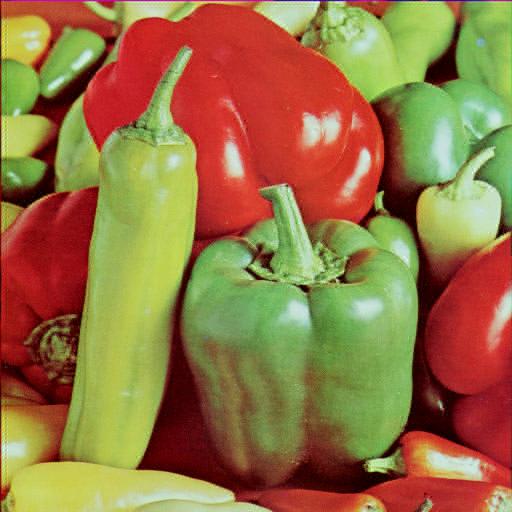}
		\caption{Color stego image}
		\label{fig:pepper color stego image}
	\end{subfigure}
	\caption{$512 \times 512$ `Pepper' color cover image and stego-image using parameter $p_1$=12, $|m|$=37.}
	\label{Figure:Color_cover_and_stego_image}
\end{figure}

Fig. \ref{Figure:Color_cover_and_stego_image} shows the subjective/ visual measure for `Pepper' color image for $p_1=12$, $|m|=37$. From this figure, we observe that the cover image and its corresponding stego-image are almost similar. Table \ref{Table:performance analysis of color image} gives the results for other measures like embedding capacity, PSNR values, mean SSIM index, NCC coefficients for the different color components, and entropy for both cover image and stego-image. {\color{myblue}We can observe from this table that the embedding capacity of our color image is approximately three times the embedding capacity of `Pepper' grayscale image for the same set of parameters. Please see columns $4$ and $8$ of Table \ref{Table:Embedding capacity for different parameter}.} This is because of the presence of three color components in the color image. Also, the PSNR values here are greater than 30 dB, and mean SSIM index \& NCC coefficients are all close to $1$, which shows that the stego-image is almost similar to its corresponding cover images. Finally, we compare the entropy of the cover image and the stego-image. We see that entropy for both these images is almost the same.

\section{Conclusions and Future Work}\label{conclusion}
We present an enhanced-embedding capacity image steganography scheme based on compressed sensing technique. 
Here, we combine three components to achieve increased embedding capacity without degrading the quality of stego-images, as well as making it resistant to steganalysis attacks. \textit{First}, we use compressed sensing to sparsify cover image block-wise and obtain its linear measurements using a matrix. We uniquely select a large number of permissible measurements. Hence, we achieve a high embedding capacity. Since the measurement matrix is a secret-key that is shared between the sender and the legitimate receiver, this adds extra security to our scheme. Also, we encrypt the secret data using the DES algorithm and then embed two bits of secret data into each permissible measurement instead of embedding one bit per measurement. \textit{Second}, we propose a technique of data extraction that is lossless and recovers our secret data entirely. \textit{Third}, we use ADMM solution of the LASSO formulation of the obtained optimization problem in the stego-image construction. The reason for selecting them is that they have broad applicability in the field of image processing, require less assumptions on the property of the objective function, have fast convergence, and are easy to implement.

We initially perform experiments on several standard grayscale images that vary in texture, and with different sets of parameters and randomly generated binary data as our secret data. For performance evaluation, we calculate embedding capacity, PSNR value, mean SSIM index, NCC coefficient, and entropy. Experiments show that our proposed CSIS achieves higher embedding capacity than existing steganography schemes that follow compression. We achieve 1.53 times more embedding capacity as compared to the most recent scheme of the similar category. PSNR values coming out of our scheme are more than 30 dB, which is considered good. Both mean SSIM index and NCC coefficients values are close to one, which shows that the cover images and their corresponding stego-images are almost similar. This similarity is further supported by the fact that we obtain approximately the same entropy value for both the cover images and their corresponding stego-images. Further, we also show the applicability of CSIS on a color image. Again, the results obtained are almost the same as that of grayscale images. However, we get approximately three times higher embedding capacity for the color image because of the presence of the three component in color images. 

In future, we plan to embed the secret data in text, audio, and video. 
{\color{myred}Other future works include extending this work for a real-time application such as hiding fingerprint data, iris data, medical information of patients, and personal signature. As mentioned in the Introduction, another line of work is embedding images inside images. Since a lot of work has been done in embedding a single image, we will focus on hiding multiple secret images and multilevel image steganography scheme.} 
\bibliographystyle{ieeetr}
 
\bibliography{mybib}


\balance

\end{document}